\documentclass{elsarticle}

\usepackage{amsmath}
\usepackage{amsfonts}
\usepackage{mathrsfs}
\usepackage{graphicx}
\usepackage{dcolumn}
\usepackage{amsmath}
\usepackage{dcolumn}
\usepackage{multirow}
\usepackage[dvips]{epsfig}
\usepackage{latexsym}
\usepackage[singlelinecheck=off,caption=false]{subfig}
\usepackage[latin1]{inputenc}
\usepackage[english]{babel}
\usepackage{verbatim}
\usepackage{setspace}
\usepackage{color}
\usepackage{epstopdf}

\journal{Phys. Lett. A}

\newcommand{\be}{\begin{equation}}\newcommand{\ee}{\end{equation}}
\newcommand{\bea}{\begin{eqnarray} }\newcommand{\eea}{\end{eqnarray}}
\newcommand{\beaa}{\begin{eqnarray} }\newcommand{\eeaa}{\end{eqnarray}}
\newcommand{\ba}{\begin{array}}\newcommand{\ea}{\end{array}}
\newcommand{\bit}{\begin{itemize}}\newcommand{\eit}{\end{itemize}}
\newcommand{\ben}{\begin{enumerate}}\newcommand{\een}{\end{enumerate}}

\def\lab{\label}

\def\rar{\rightarrow}

\def\al{\alpha}

\def\1{{_{1}}}\def\2{{_{2}}}

\def\lsim{\hbox{ \raise.35ex\rlap{$<$}\lower.6ex\hbox{$\sim$}\ }}
\def\gsim{\hbox{ \raise.35ex\rlap{$>$}\lower.6ex\hbox{$\sim$}\ }}

\def\lab{\label}

\def\rar{\rightarrow}

\def\al{\alpha}

\begin{document}

\begin{frontmatter}

\title{On the coherent behavior of pancreatic beta cell clusters}
\author[a1]{Alessandro Loppini}
\ead{a.loppini@unicampus.it}
\author[a2]{Antonio Capolupo\corref{cor1}}
\ead{capolupo@sa.infn.it}
\author[a1,a3]{Christian Cherubini}
\ead{c.cherubini@unicampus.it}
\author[a1]{Alessio Gizzi}
\ead{a.gizzi@unicampus.it}
\author[a4]{Marta Bertolaso}
\ead{m.bertolaso@unicampus.it}
\author[a1,a3]{Simonetta Filippi}
\ead{s.filippi@unicampus.it}
\author[a2]{Giuseppe Vitiello}
\ead{vitiello@sa.infn.it}

\cortext[cor1]{Corresponding author: \\
phone: \\
address: }
\address[a1]{Nonlinear Physics and Mathematical Modeling Lab,\\
University Campus Bio-Medico, Via A. del Portillo 21, I-00128 Rome, Italy.}
\address[a2]{Physics Department,  University of Salerno, Baronissi (SA) - 84081 Italy.}
\address[a3]{International Center for Relativistic Astrophysics.}
\address[a4]{Istituto di Filosofia dell'Agire Scientifico e Tecnologico (FAST),\\
University Campus Bio-Medico, Via A. del Portillo 21, I-00128 Rome, Italy.}

\begin{abstract}
Beta cells in pancreas represent an example of coupled biological oscillators which via communication pathways, are able to synchronize their electrical activity, giving rise to pulsatile insulin release.
In this work we numerically analyze scale free self-similarity features of membrane voltage signal power density spectrum, through a stochastic dynamical model for beta cells in the islets of Langerhans fine tuned on mouse experimental data. Adopting the algebraic approach of coherent state formalism, we show how coherent molecular domains can arise from proper functional conditions leading to a parallelism with "phase transition" phenomena of field theory.
\end{abstract}


\end{frontmatter}

\section{Introduction}
\label{sec:Intro}
Islets of Langerhans in the pancreas are ellipsoidal clusters of excitable endocrine cells that ensure blood glucose homeostasis. Alpha, beta, delta and PP cells form this particular structure. Specifically beta cells are able to lower glycemic level by releasing insulin~\cite{Ash}. In rodents, these cells are clustered in the central core of the islet surrounded by peripheral alpha cells, and are coupled through specific connections, the gap junctions~\cite{Briss,Stein,Orc,Mic}. In response to glucose uptake from extracellular space, beta cells within the islet modify their membrane potential, exhibiting slow oscillations with superimposed action potentials (bursting activity)~\cite{Ash}. This characteristic behavior leads to oscillations of the intracellular calcium concentration that triggers pulsatile insulin release~\cite{Roc,Gilon}. On the other hand, isolated beta cells show an irregular spiking activity in response to glucose stimuli. Experimental measurements show moreover that beta cells electrical activity is synchronized over the islet~\cite{Roc,San,Ped}. Such an observed feature highlights that a coherent intercellular correlation extending over the islet's volume can be established under proper functional conditions and glucose concentrations, providing the motivation for the study presented in this paper. We indeed analyze such a behavioral property of the cells in the frame of a Hodgkin-Huxley~\cite{Hodgk} type model also considering the kinetics of the stochastic K-Ca channels~\cite{Sher} (calcium-dependent potassium channels; here Ca stands for the ion Ca$^{2+}$). We thus focus our study on the dynamics of the membrane potential in variable scale clusters in connection with intracellular calcium concentration and the activation of potassium and calcium channels. Our numerical simulations agree with the observed possibility of formation of coherent molecular domains and show how this depends on glucose concentrations and on the islet size. In particular, we find that the power density spectrum (PDS) of membrane voltage exhibits scale free self-similarity features with respect to frequency occurring with different self-similarity dimension in different frequency intervals and show how such a feature is indeed evidence of coherent molecular dynamics. In our discussion, we use the algebraic approach of the coherent state formalism through which the isomorphism is shown to exist between the linear fit of the log-log plot of PDS/frequency and the squeezed coherent state algebraic structure. The plan of the paper is the following. The mathematical modeling is introduced in Section~\ref{sec:Math}, while numerical simulations and results are presented in Section~\ref{sec:sim}. The notions of (fractal) self-similarity dimension and squeezed coherent states are given in Section~\ref{sec:self-sim}, where the algebraic isomorphism between self-similarity and coherent states is also discussed. Section~\ref{sec:concl} is devoted to conclusions.

\section{Mathematical Modeling}
\label{sec:Math}
\begin{figure}[h]
\centering
\subfloat[]{\includegraphics[scale=0.3]{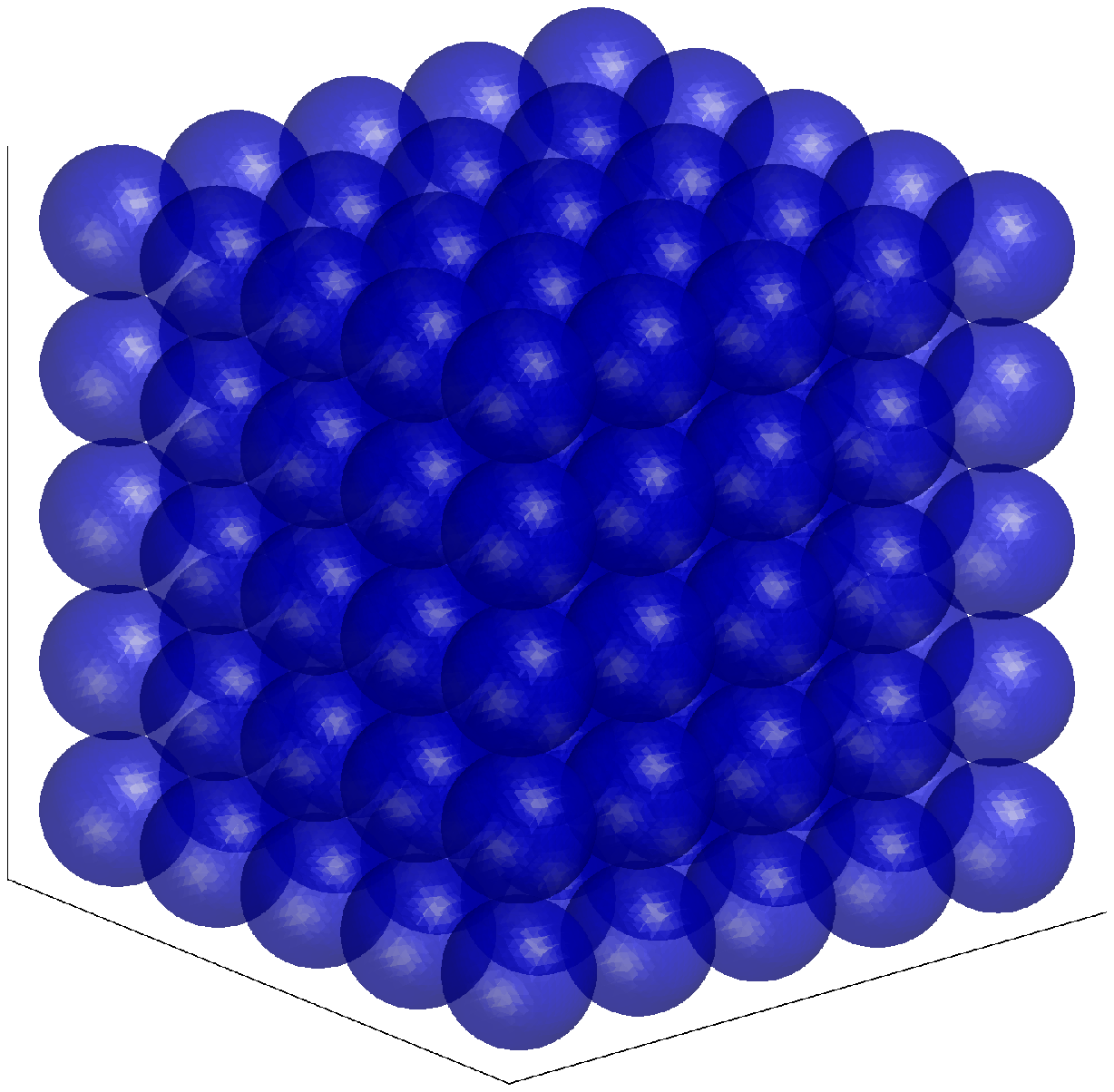}
\label{fig:clust-a}}
\subfloat[]{\includegraphics[scale=0.3]{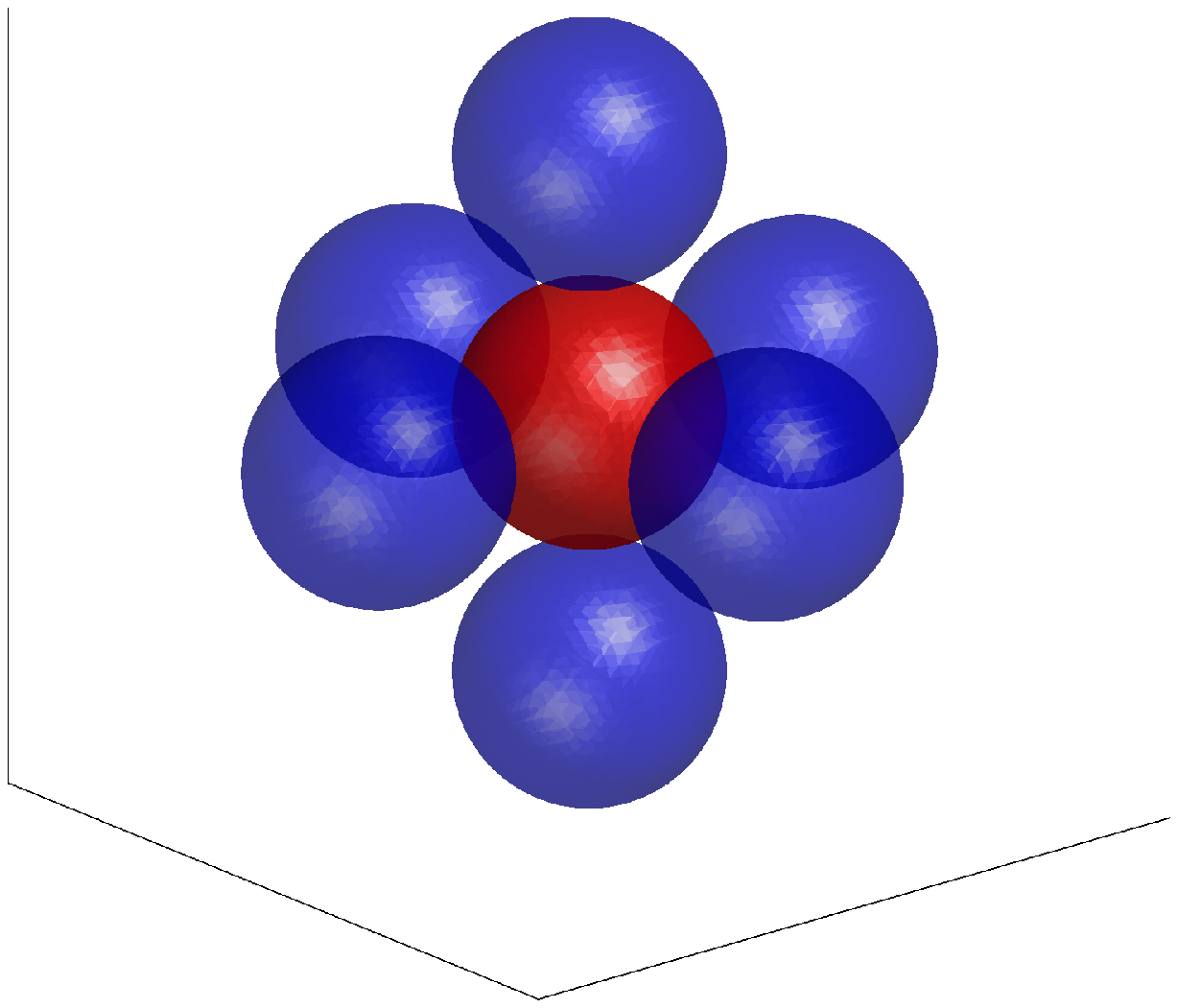}
\label{fig:clust-b}}
\caption{3D view of a $5 \times 5 \times 5$ cluster (a) and Von Neumann neighborhood (b) for a central cell, plotted in red.}
\label{fig:cluster}
\end{figure}

An extended version of the stochastic multi-cell SRK model \cite{She}, based on mouse electrophysiological data, was adopted to study bursting activity of beta cells cubic clusters of variable size for different glycemic levels. Such an Hodgkin-Huxley type model permits to reproduce isolated and coupled dynamics of beta cells.
This is because of the stochastic formulation of the high conductance K-Ca channel gating. Single gating events of this specific type of channel can lead the cell to an active or a silent state, making possible to observe a spiking activity in an isolated cell. Thanks to "channel sharing", coupled cells are able to share the entire population of K-Ca channels \cite{She,Sher}, overcoming noise and giving rise to bursting behavior, a characteristic electrical pattern observed experimentally in beta cells within the islet.
The model's equations for the i-th cell are:

\begin{eqnarray*}
  \label{eq:volt}
  C_m\frac{dV}{dt} &=& -I_{ion}-\bar{g}_{K-Ca}p(V-V_K)-g_c\sum_{j \in \Omega}(V-V_j)
  \\
  \label{eq:vargate}
  \frac{dn}{dt} &=& \lambda\left[\frac{n_{\infty}-n}{\tau_n(V)}\right]
  \\
  \frac{dCa}{dt} &=& f\left[-\alpha I_{Ca}-k_{Ca}Ca\right]
  \\
  \langle p \rangle &=& \frac{Ca}{K_d+Ca}
  \\
  I_{ion} &=& I_K+I_{Ca}=\bar{g}_Kn(V)(V-V_K)+\bar{g}_{Ca}m_\infty(V)h(V)(V-V_{Ca})
  \\
  m_\infty(V)&=&  \frac{1}{1+\exp[(V_m-V)/S_m]}
  \\
  h(V) &=& \frac{1}{1+\exp[(V-V_h)/S_h]}
  \\
  n_\infty(V) &=&  \frac{1}{1+\exp[(V_n-V)/S_n]}
  \\
  \tau_n(V) &=& \frac{c}{\exp[(V-\bar{V})/a]+\exp[(V-\bar{V})/b]} \,,
\end{eqnarray*}
where the dynamical variables $V,\,n,\,Ca$ have to be understood as $V_i,\,n_i,\,Ca_i$.
The ODE system models the dynamics of the membrane potential $V$, the potassium channel activation level $n$ and the intracellular calcium concentration $Ca$. $C_m$ represents the membrane capacitance; $n_\infty(V)$ and $m_\infty(V)$ are the steady state activation functions of potassium and calcium channels, respectively; $h(V)$ is the inactivation curve of calcium channels; $\tau_n(V)$ is the voltage dependent time constant of potassium channels fine tuned by the parameter $\lambda$; $\bar{g}_K$, $\bar{g}_{Ca}$ and $\bar{g}_{K-Ca}$ represent the whole cell conductances of potassium, calcium and potassium-calcium dependent ionic channels, respectively; $g_c$ is the strength (conductance) of coupling between two adjacent cells due to gap junctions, in a 3D Von Neumann neighborhood ($\Omega$) of the cell modified to take account of different communication rules on the boundary region (in Fig.~\ref{fig:cluster} is shown a $5 \times 5 \times 5$ cluster and the neighborhood considered for a central cell); $p$ is the fraction of open K-Ca channels whose transition events are obtained as evolution of a stochastic process described in the following; $V_K$ and $V_{Ca}$ are the potassium and calcium equilibrium potentials; $f$ is a fixed parameter which slows down intracellular calcium dynamics; the factor $\alpha$ converts current units in concentration units; $k_{Ca}$ represents the rate at which calcium is pumped out from cytosol to extracellular space; $K_d$ is a factor that depends on K-Ca channels kinetics.

\subsection{Stochastic gating of K-Ca channels}
Transition events of the K-Ca channels were computed with the use of a two state Markov process for each ionic channel.
Following \cite{Sher}, the following kinetics was adopted:
$$
\begin{array}{ccc}
 & 1/\tau_c & \\
 \mathcal{C} & \rightleftharpoons & \mathcal{O} \\
 & 1/\tau_o &
\end{array} \, ,\quad
\tau_o=\tau_c\frac{Ca_i}{K_d} \,.
$$

Here $\tau_o$ and $\tau_c$ are the mean opening and closing times.
Keeping $\tau_o$ fixed, $\tau_c$ varies as a function of intracellular calcium.
The probabilities that a channel in a specific state makes a transition in a fixed time window are given by:
$$
\frac{\Delta t}{\tau_c}= \mathrm{Prob} \left\{ s=\mathcal{O}, t+\Delta t \mid s=\mathcal{C}, t \right\}\,,\quad
\frac{\Delta t}{\tau_o}= \mathrm{Prob} \left\{ s=\mathcal{C}, t+\Delta t \mid s=\mathcal{O}, t \right\}\,;
$$
where $s\in\{C,O\}$ is a stochastic variable and $\Delta t$ is the considered time step.\cite{Fall}
Considering 600 channels of K-Ca type per cell, each two-state process was resolved with the use of a Monte Carlo simulation, computing the fraction of open channels at every integration time step of the model equations.

\subsection{Glucose feedback}
As in ref.\,\cite{Portu}, glucose feedback was modeled tuning the calcium removal rate parameter in order to reproduce experimentally observed beta cells activity at specific blood glucose concentrations $[G]$. At about $[G]=5.5\,mM$ and $[G]=16.6\,mM$, a silent-bursting and a bursting-continuous spiking transition, respectively, can be observed in beta cells membrane potential. These behaviors can be obtained in the model setting $k_{Ca}=0.02\,ms^{-1}$ and $k_{Ca}=0.09\,ms^{-1}$. Considering these observations, a simple linear function was adopted to achieve the feedback:
\begin{equation*}
\label{eq:kca}
k_{Ca}=A [G] - B  \quad \mathrm{for} \quad  [G] \geq 2.33\,mM \, ,
\end{equation*}
where $A=6.3\cdot10^{-3}\,ms^{-1}\,mM^{-1}$, $B=0.0147\,ms^{-1}$ and $[G]$ is the glucose concentration.

\section{Numerical simulations}
\label{sec:sim}
A fourth order Runge-Kutta method with a fixed time step of $0.1\,ms$ was adopted to solve the ODE system. The global model was implemented in a C++ code. Cells' voltage membrane signals obtained from simulations were analyzed with the Matlab\textsuperscript{\textregistered} software, computing signals spectra with the use of the FFT algorithm here implemented. Our analysis shows how the activity of beta cells depends on the glucose bath concentration $[G]$ and on the islet size (cluster population).

\paragraph{\it Dependence of the beta cell activity on glucose bath concentration}

\begin{figure}[h]
\centering
\subfloat[]{\includegraphics[scale=0.3]{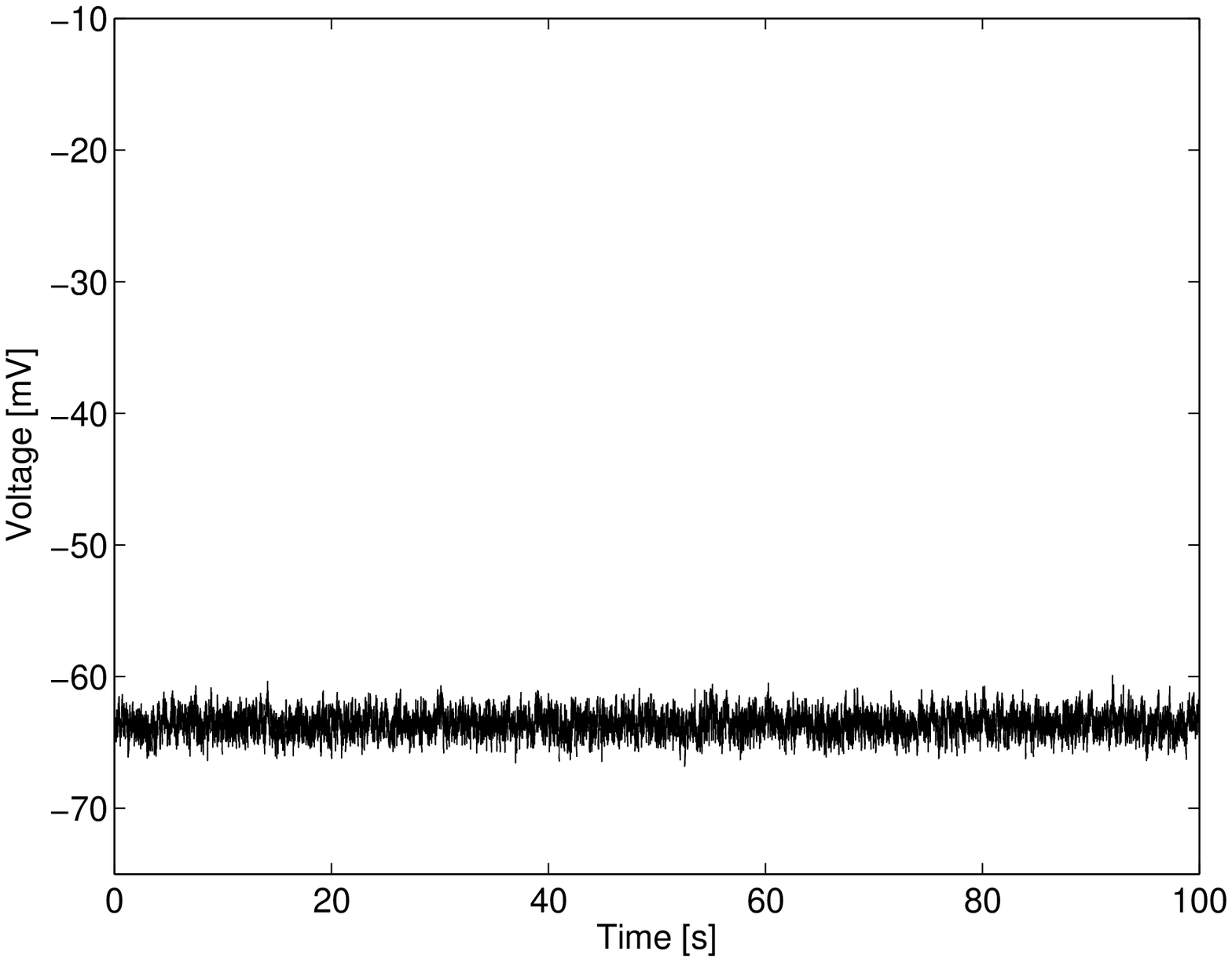}
\label{fig:pot-a}} \hspace{5mm}
\subfloat[]{\includegraphics[scale=0.3]{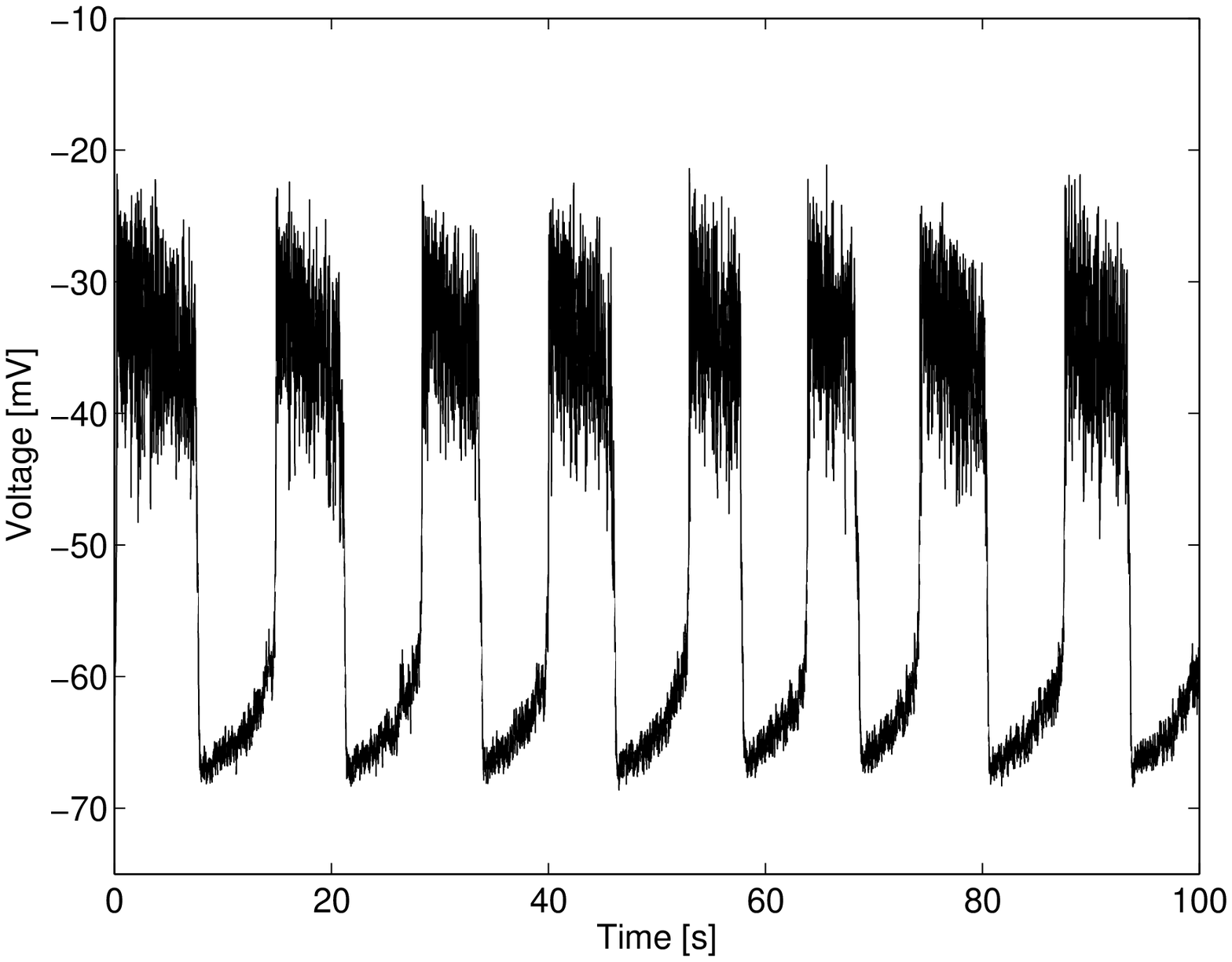}
\label{fig:pot-b}} \hspace{5mm} \\
\subfloat[]{\includegraphics[scale=0.3]{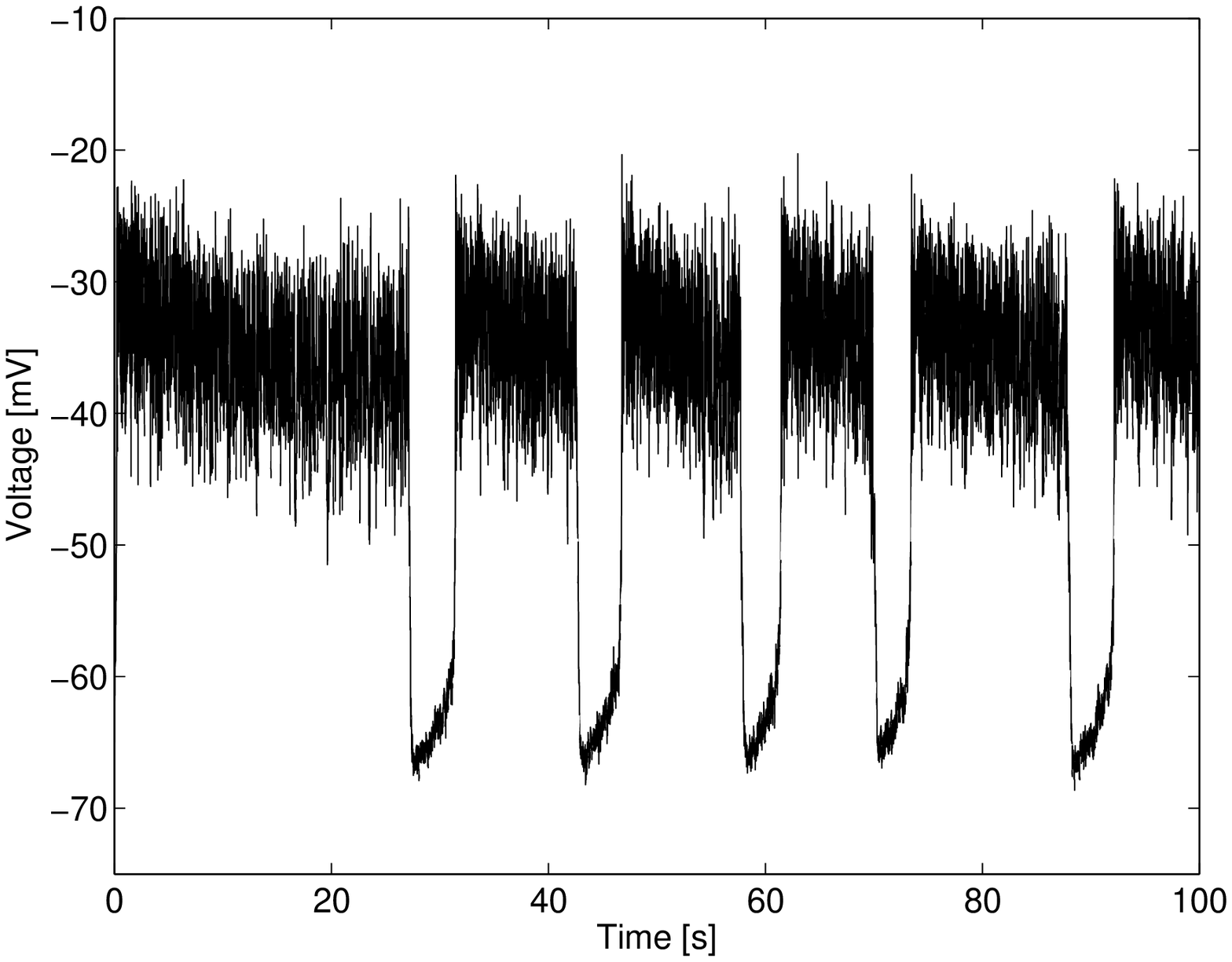}
\label{fig:pot-c}} \hspace{5mm}
\subfloat[]{\includegraphics[scale=0.3]{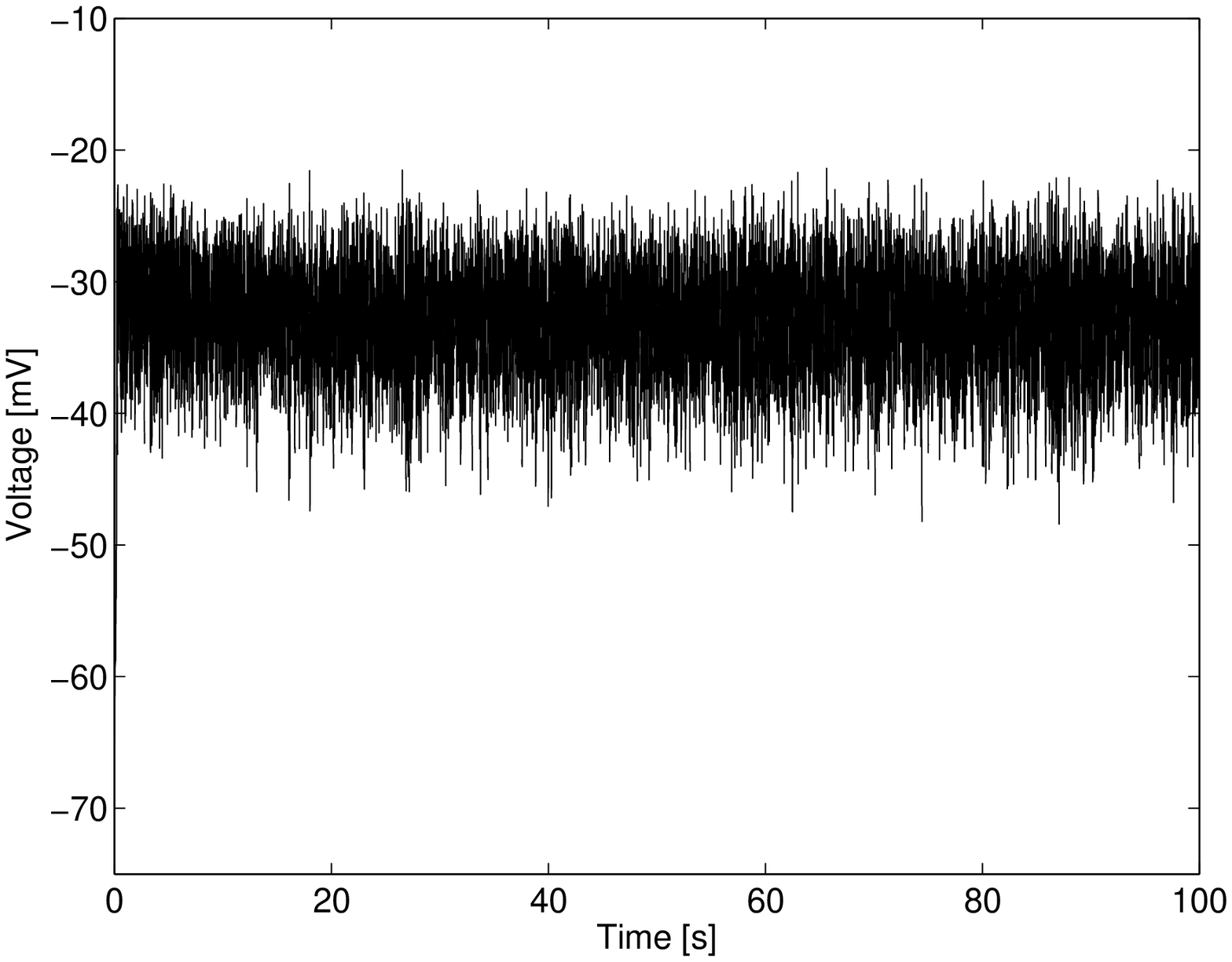}
\label{fig:pot-d}}
\caption{Membrane potential signals of a representative cell in a $5 \times 5 \times 5$ cluster, stimulated by different glucose concentrations: (a) $[G]=4.7\,mM$; (b) $[G]=9.5\,mM$; (c) $[G]=12.6\,mM$; (d) $[G]=16.6\,mM$.}
\label{fig:pot}
\end{figure}

\begin{figure}[h]
\centering
\includegraphics[scale=0.4]{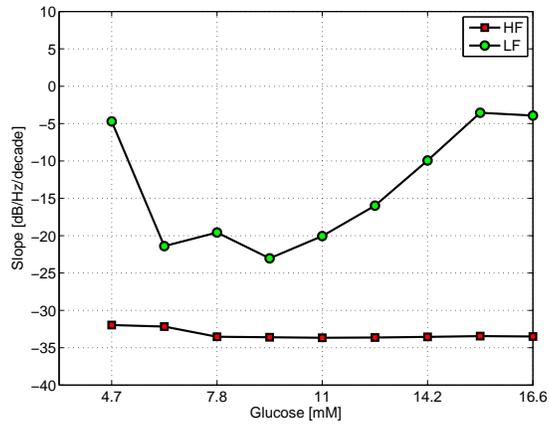}
\caption{Computed slopes at low (green circles) and high (red squares) frequencies for increasing glucose concentrations.}
\label{fig:slope-g}
\end{figure}

In order to investigate a coherent activity of beta cells in the islet and its possible correlation with glucose stimuli, we initially modeled a 125 cells cubic cluster in glucose baths of different concentrations.
Specifically, the cluster response was studied for glycemic levels in the range $[G]=4.7\,mM-16.6\,mM$, analyzing a representative cell of the network. In Fig.~\ref{fig:pot} the computed membrane potentials in time for $[G]=4.7,\,9.5,\,12.6,\,16.6\,mM$ are shown (Fig.~\ref{fig:pot}a, b, c, and d, respectively). As can be seen, the model correctly reproduces silent, bursting and continuous bursting behavior experimentally observed at increasing glucose levels. Three characteristic time scales can be noticed: the bursting slow wave with a period of tens of seconds, the spiking activity of the order of hundreds of milliseconds, and an infinitely low time scale due to stochastic channel gating.
The power density spectrum (PDS) of voltage signals was computed to highlight scaling features of the signals as an observable phenomena of coherent activity in the cluster. In Fig.~\ref{fig:tr} the PDS of signals reported in Fig.~\ref{fig:pot} are plotted in a log scale.
Scaling properties were then analyzed computing log-log slopes of linear trends observed in PDS. Performed analysis have shown that a representative cell of the cluster, stimulated by typical postprandial glucose concentration ($[G]\simeq9.5\,mM$, Fig.~\ref{fig:tr-b}), presents two different linear trends in the log-log plot of PDS, at low and high frequency ranges, respectively, with an intermediate transition region (highlighted in blue in the PDS plots).
The slopes in the log-log plots were then computed with a linear fitting of data in these two ranges of frequencies (red and green lines in the PDS plots represent the computed slopes), at different glucose levels. Results obtained are reported in Fig.~\ref{fig:slope-g}. As can be seen, the slope at high frequencies (HF) is $S_{HF}\simeq-33\,dB/Hz\,\mathrm{per\,decade}$ and seems to be independent on glucose level. The slope at low frequencies (LF) instead is nearly absent in the clustered cells stimulated by both subthreshold and very high glucose concentrations, while can be observed at intermediate glucose levels that ensure regular bursting, where $S_{LF}\simeq-20\,dB/Hz\,\mathrm{per\,decade}$ at postprandial glycemic level.

\paragraph{\it Dependence of the beta-cell activity on cluster population}

\begin{figure}[h]
\centering
\subfloat[]{\includegraphics[scale=0.3]{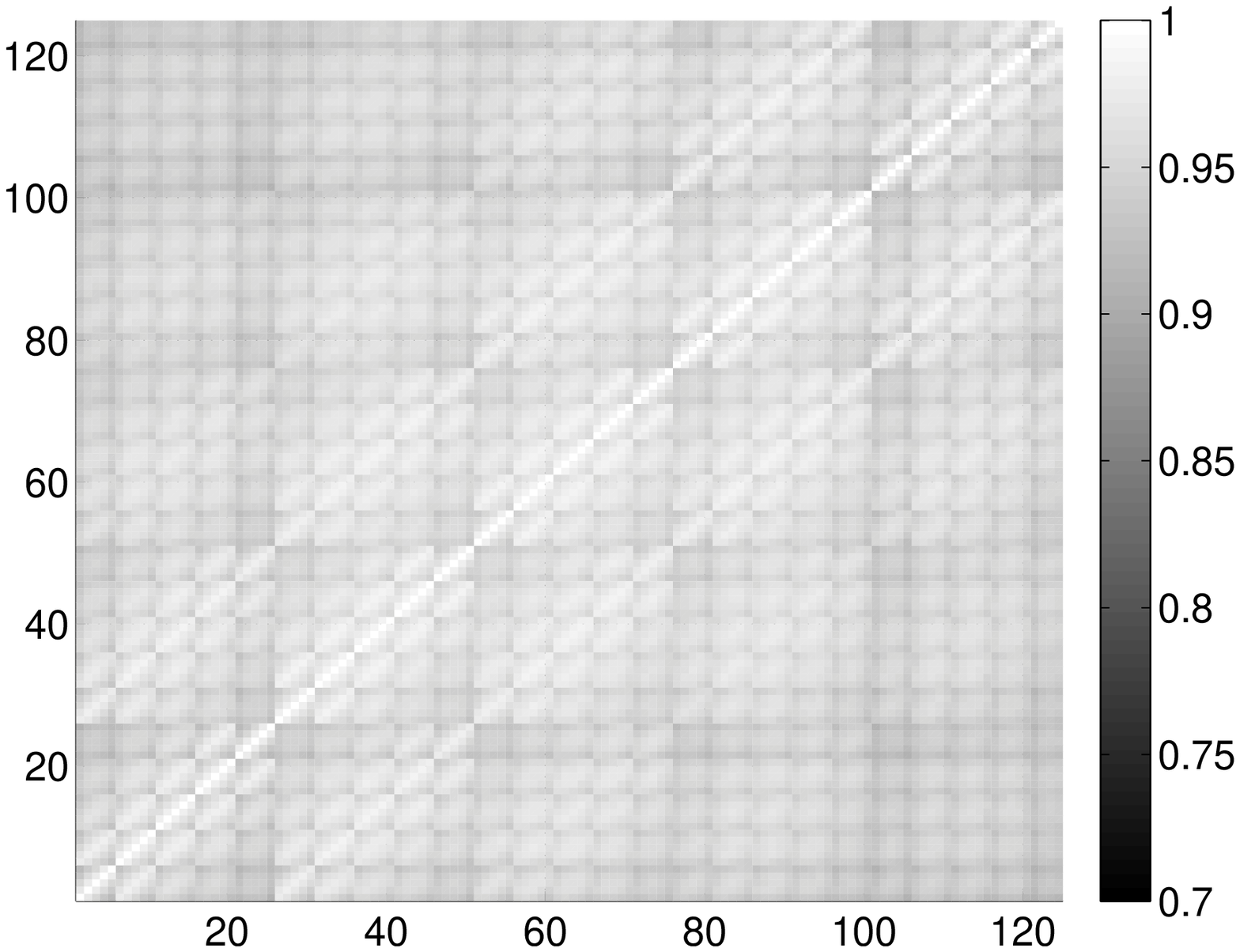}
\label{fig:corr-a}} \hspace{5mm}
\subfloat[]{\includegraphics[scale=0.3]{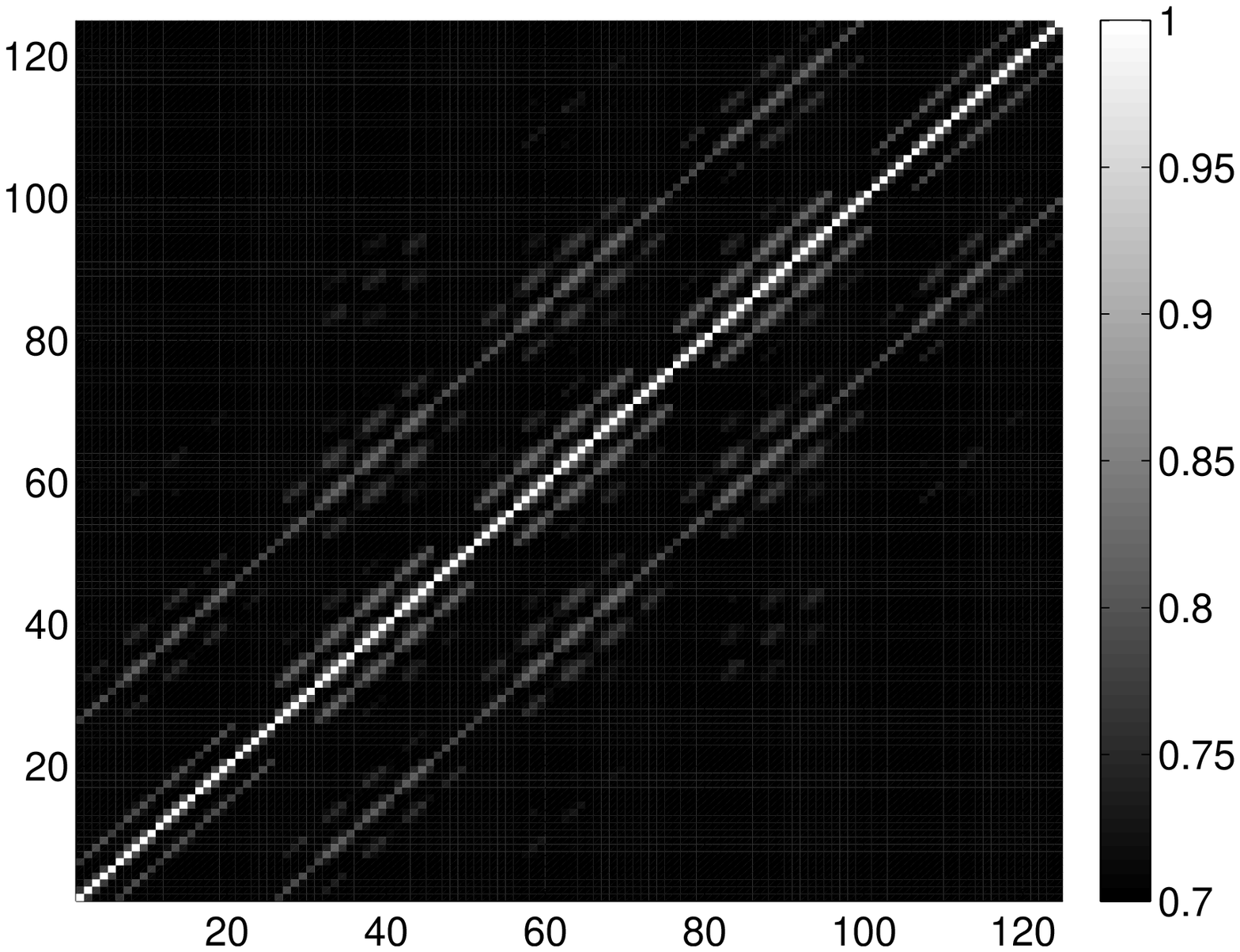}
\label{fig:corr-b}}
\caption{Surface plots of the correlation matrix for (a) postprandial and (b) subthreshold glucose concentration in a $5 \times 5 \times 5$ beta cells cluster. The labeling of the axes has to be understood as a sequential numbering of the cubic cluster: 1-25 first cluster plane; 26-50 second plane; 51-75 third plane; 76-100 fourth plane; 101-125 fifth plane.}
\label{fig:corr}
\end{figure}

With the aim to support the coherence hypothesis, the Pearson's correlation matrix was obtained from the data, treating membrane potential time series of the modeled cells as multiple observations from different stochastic variables. In Fig.~\ref{fig:corr}, the surface plots of the correlation matrix for the 125 cells cubic cluster in subthreshold and postprandial glucose concentrations are shown.
As can be noticed, the correlation indexes of cells' voltage signals are quite high (all above 0.9) for suitable glucose levels (Fig.~\ref{fig:corr-a}), both for neighboring cells and for different cellular planes of the cluster; considering low glucose concentration instead, correlation indexes get lower with minimum values of about 0.4 (Fig.~\ref{fig:corr-b}).

\begin{figure}[h]
\centering
\includegraphics[scale=0.4]{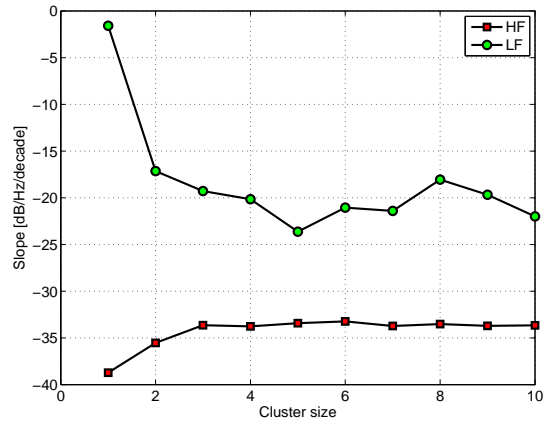}
\caption{Computed slopes at low (green circles) and high (red squares) frequencies for increasing cluster size. The value $n$ on the x axis has to be understood as $n \times n \times n$.}
\label{fig:slope-s}
\end{figure}

Because of channel noise, a minimum size of the cluster  (minimum cell population) is required to achieve robust bursting. With the aim to study cell population effect on slopes in the log-log PDS plots, we modeled clusters of increasing sizes starting from the single isolated cell up to the 1000 cells case, stimulated by a postprandial glucose concentration. In Fig.~\ref{fig:trasf} the PDS plots of voltage signals for a representative cell of each cluster are reported. From results in Fig.~\ref{fig:slope-s}, the slope at high frequencies seems to be also independent on cluster size with $S_{HF}\simeq-33\,dB/Hz\,\mathrm{per\,decade}$ for clusters bigger than about ten cells; slightly lower values were obtained for the isolated cell and for the $2 \times 2 \times 2$ cluster.
This threshold value for the cluster size can be noticed also in $S_{LF}$ that is almost zero in the single cell, while increase in modulus starting from the $2 \times 2 \times 2$ cluster.

In order to test slope invariance under perturbations, $\delta V$, of the single cell membrane potential $V$, model equations were solved considering the new variable  (linear transformation) $\widetilde{V}=V+V_t$, with $V_t \equiv \delta V$ a constant perturbation parameter. Clusters containing 1, 27, 64 and 125 cells stimulated by postprandial glucose concentration were modeled analyzing log-log PDS plots of $\widetilde{V}$ for different values of $V_t$. Results are summarized in Tab.~\ref{tab:trasf}.
As it can be seen, apart the case of 1 cell, there are very little changes in the values of slopes, which signals the slope invariance
under the $V_t$ perturbation.

\section{Self-similarity analysis}
\label{sec:self-sim}
\begin{figure}[h]
\centering
\subfloat[]{\includegraphics[scale=0.3]{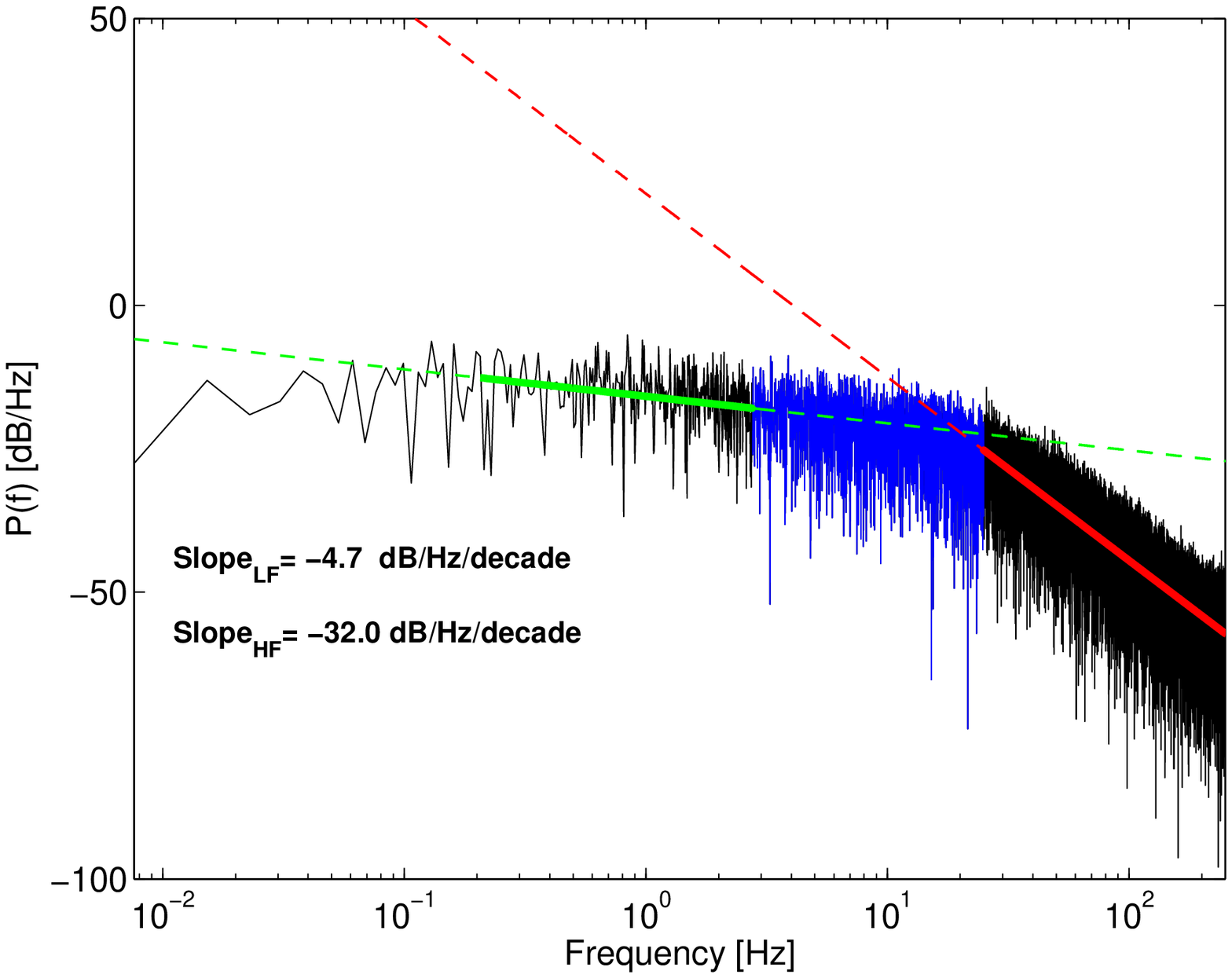}
\label{fig:tr-a}} \hspace{5mm}
\subfloat[]{\includegraphics[scale=0.3]{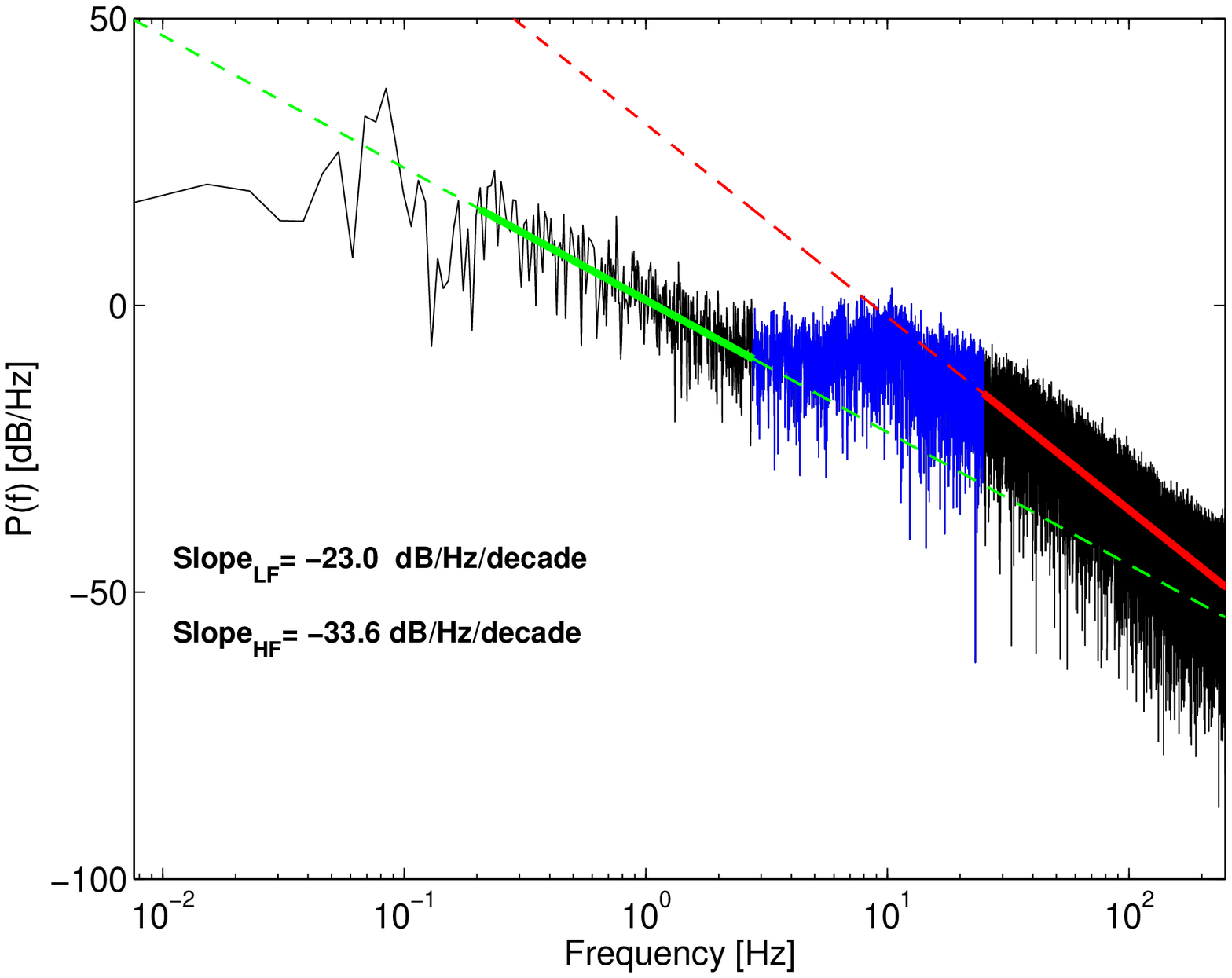}
\label{fig:tr-b}} \hspace{5mm}\\
\subfloat[]{\includegraphics[scale=0.3]{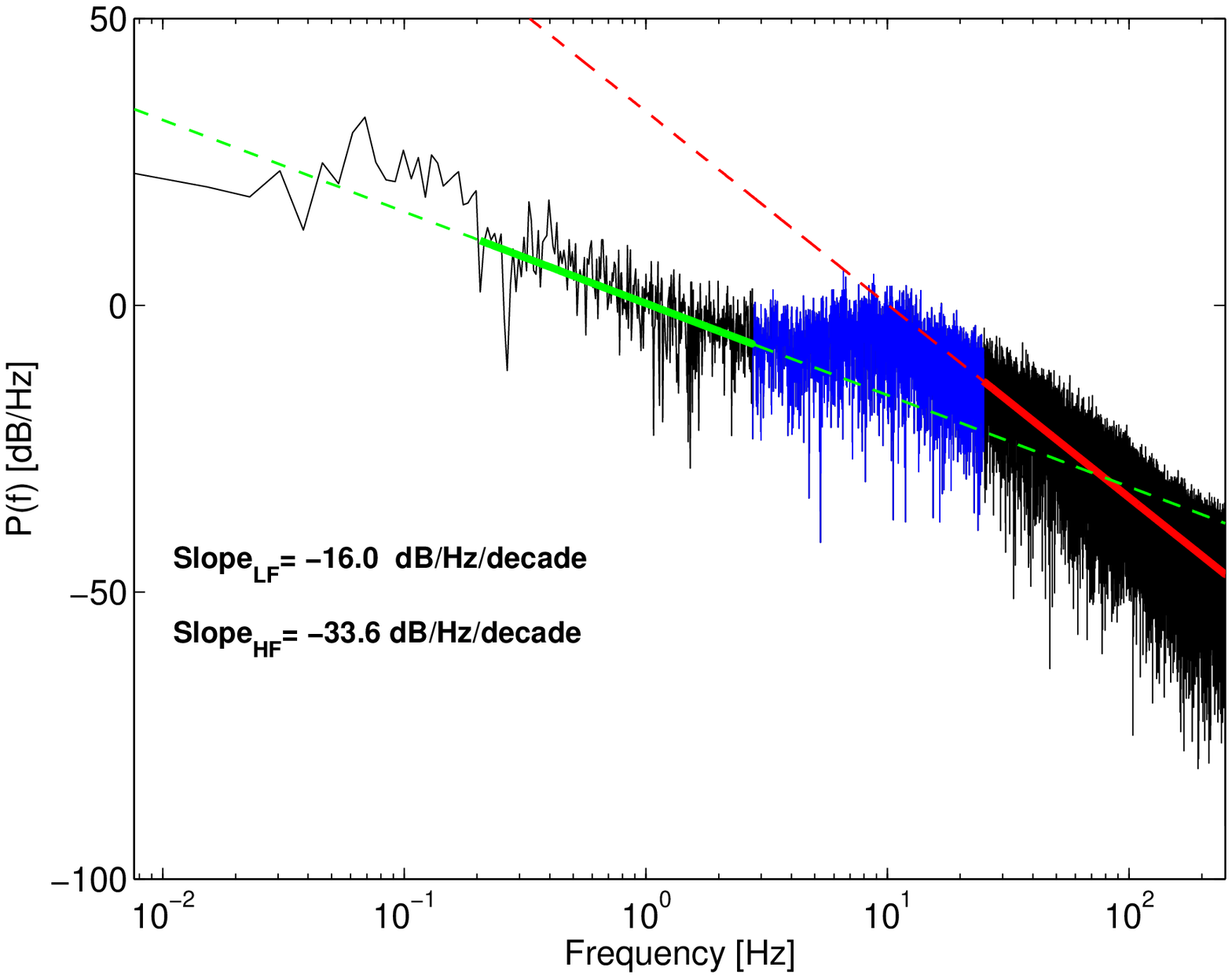}
\label{fig:tr-c}} \hspace{5mm}
\subfloat[]{\includegraphics[scale=0.3]{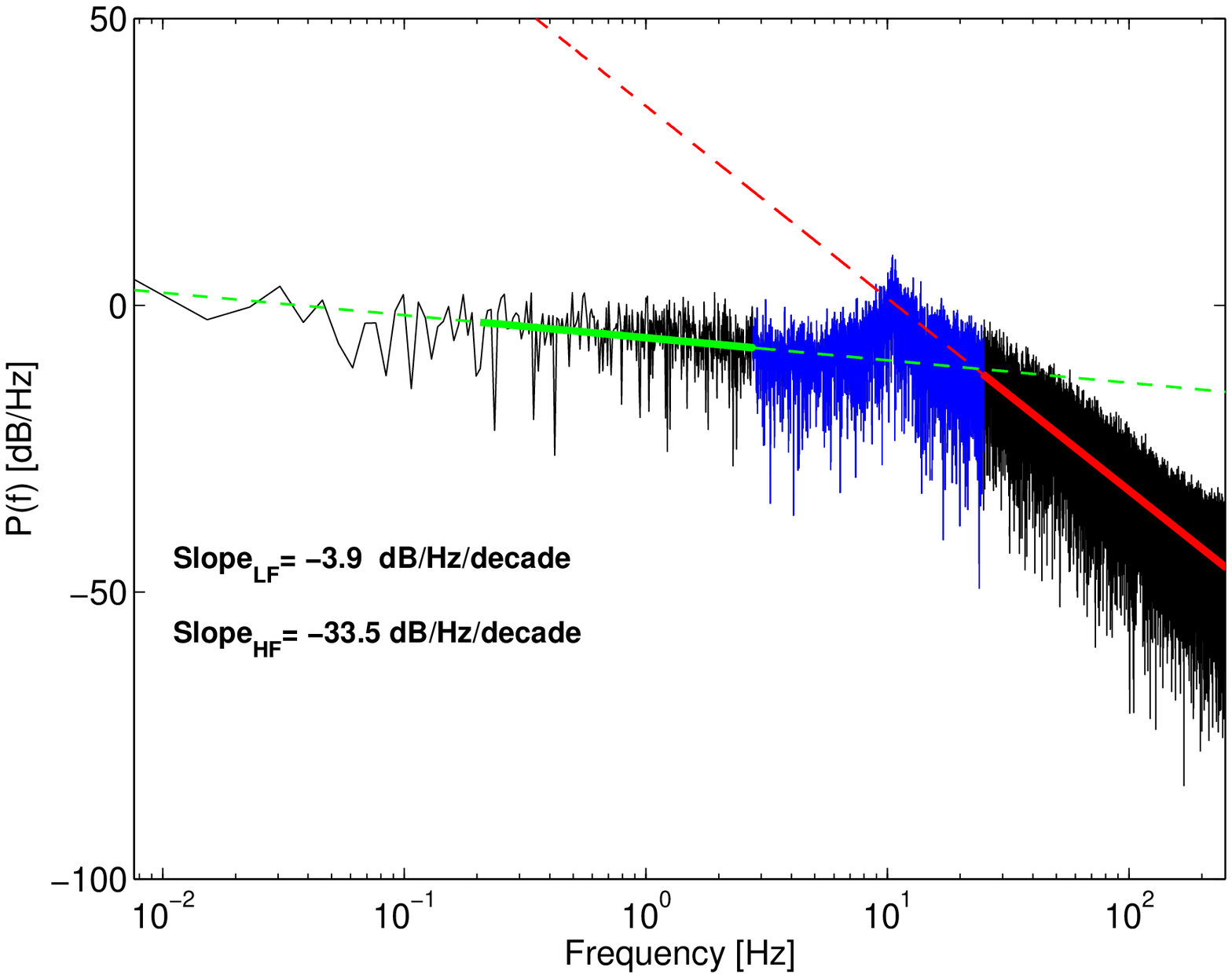}
\label{fig:tr-d}}
\caption{Log-log power density spectra of membrane potential signals of a representative cell in a $5 \times 5 \times 5$ cluster, stimulated by different glucose concentrations: (a) $[G] = 4.7\,mM$; (b) $[G] = 9.5\,mM$; (c) $[G] = 12.6\,mM$; (d) $[G] = 16.6\,mM$.
The slope at low frequencies ($S_{LF}$) is highlighted in green, the slope at high frequencies ($S_{HF}$) in red. Continuous lines segments highlight the PDS points used for the linear fitting; dotted lines segments are the extrapolation of the linear estimation. The transition region between the two linear zones is highlighted in blue.}
\label{fig:tr}
\end{figure}

\begin{figure}[h]
\centering
\subfloat[]{\includegraphics[scale=0.2]{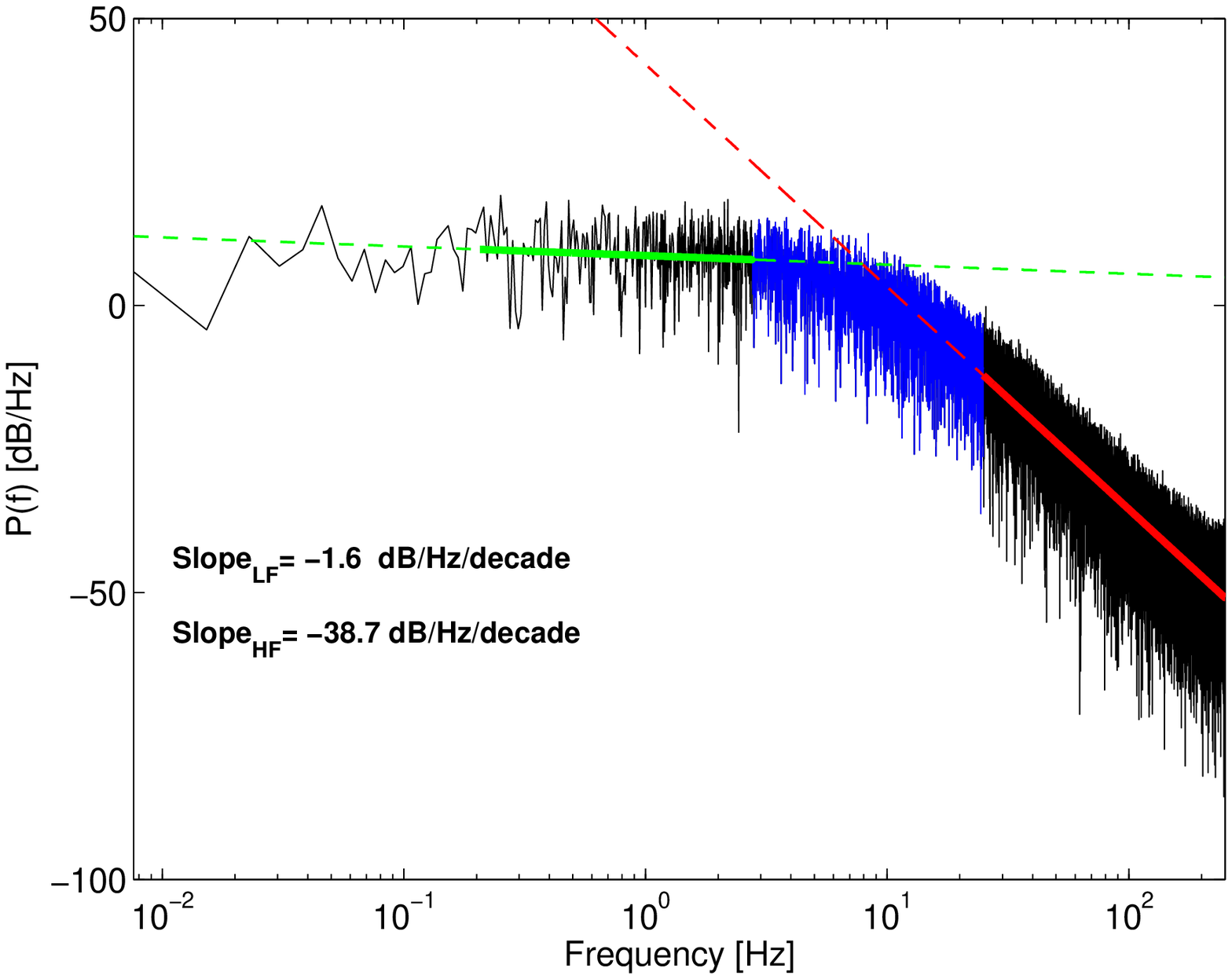}
\label{fig:trasf-a}} \hspace{2mm}
\subfloat[]{\includegraphics[scale=0.2]{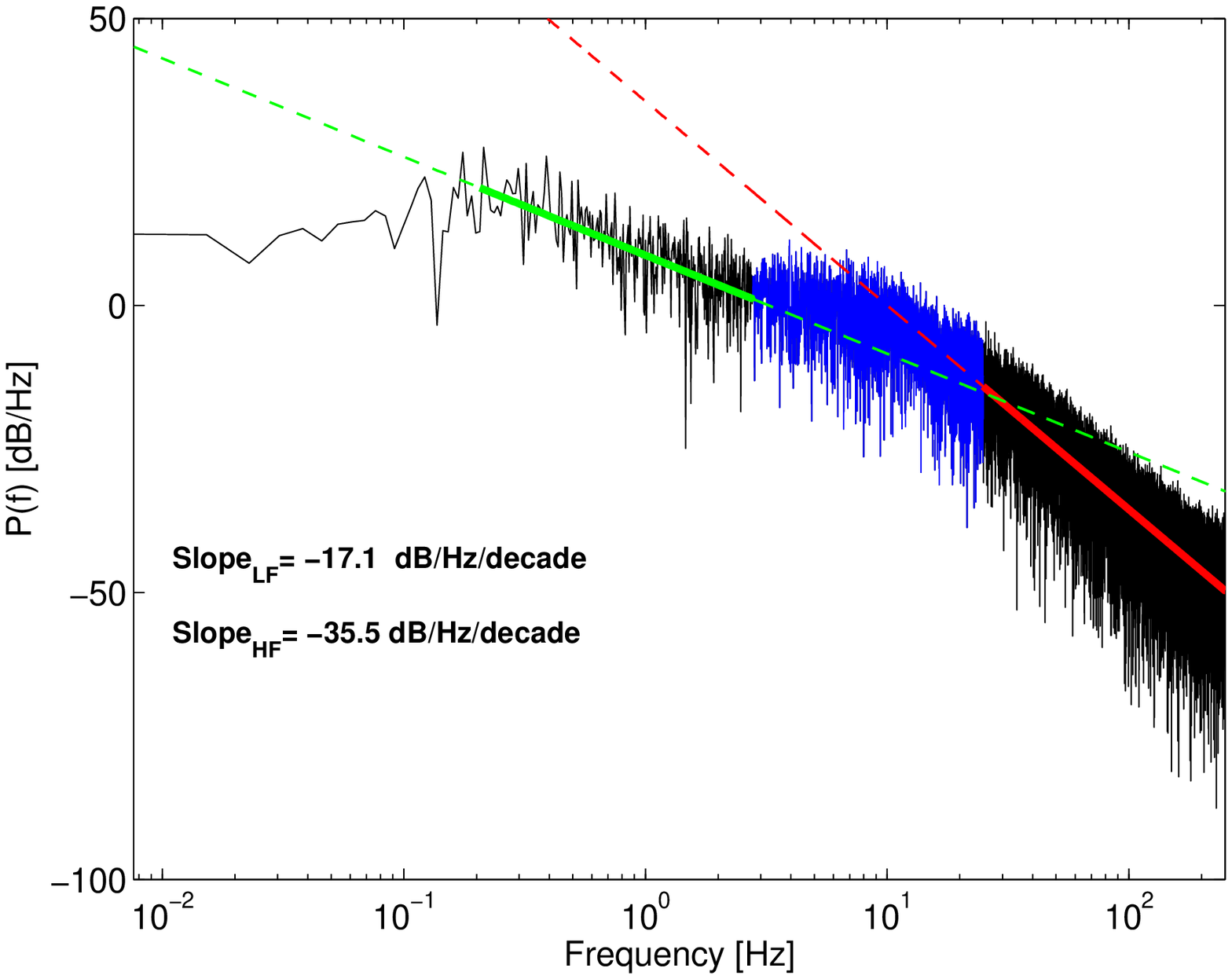}
\label{fig:trasf-b}} \hspace{2mm}
\subfloat[]{\includegraphics[scale=0.2]{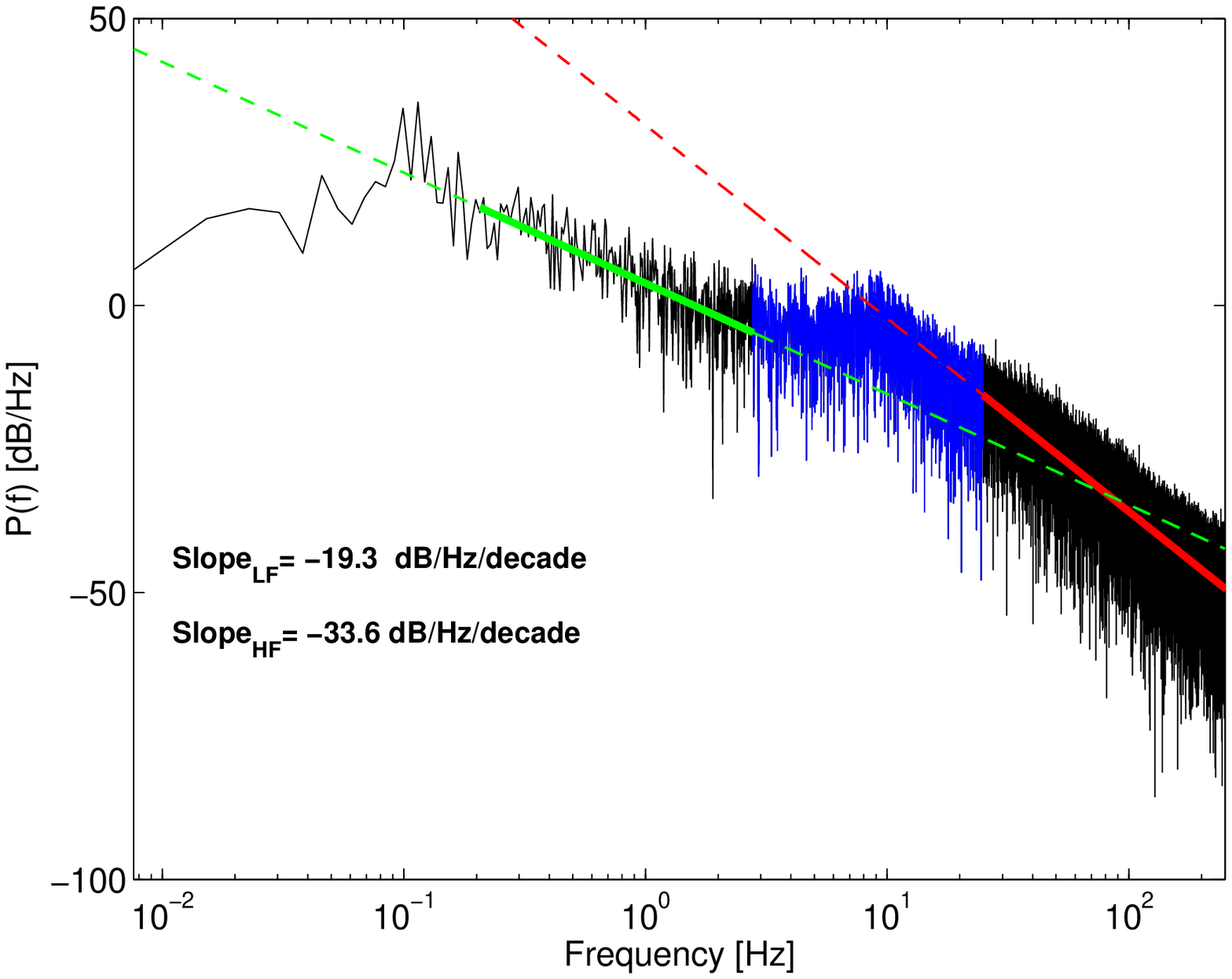}
\label{fig:trasf-c}} \\
\subfloat[]{\includegraphics[scale=0.2]{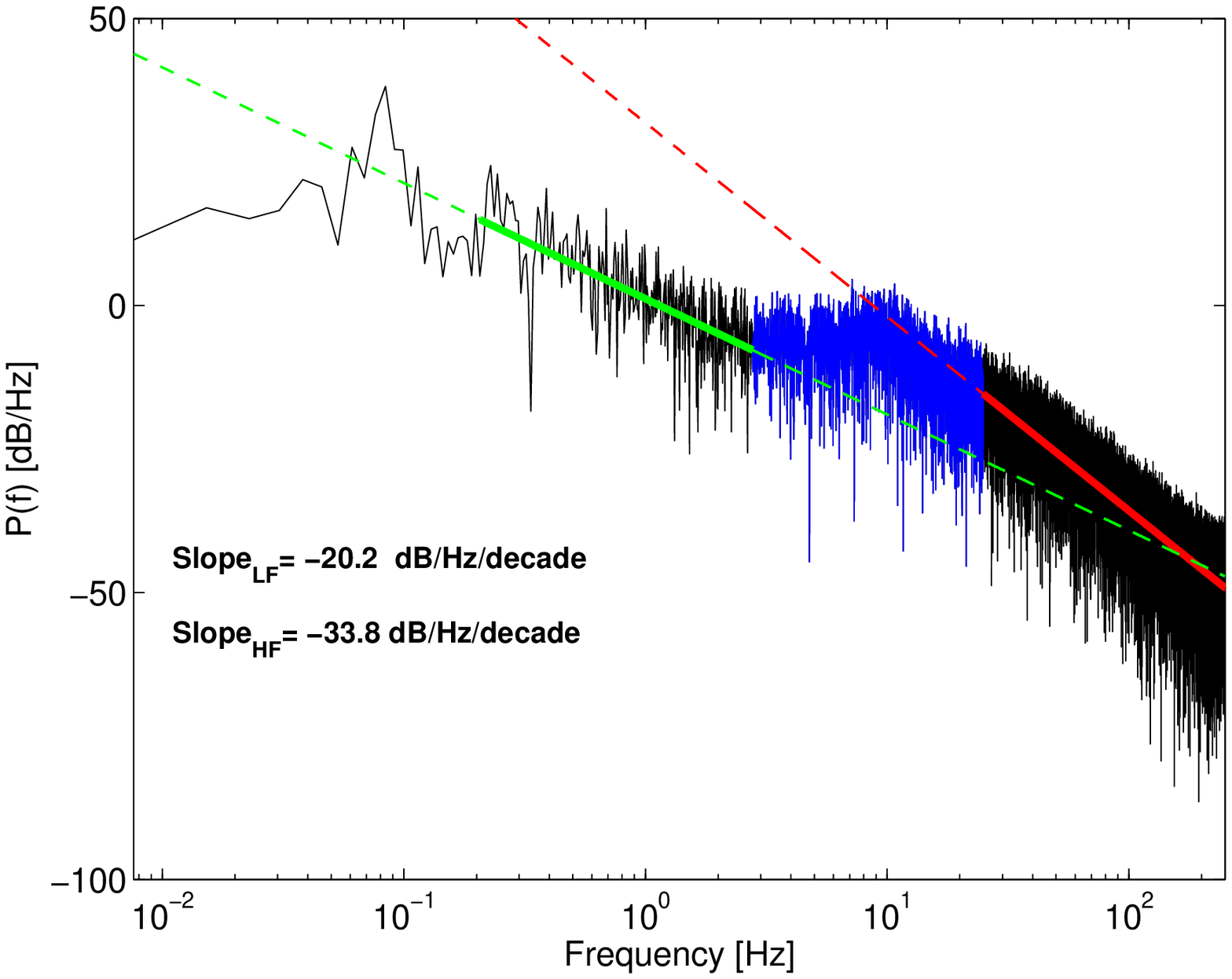}
\label{fig:trasf-d}} \hspace{2mm}
\subfloat[]{\includegraphics[scale=0.2]{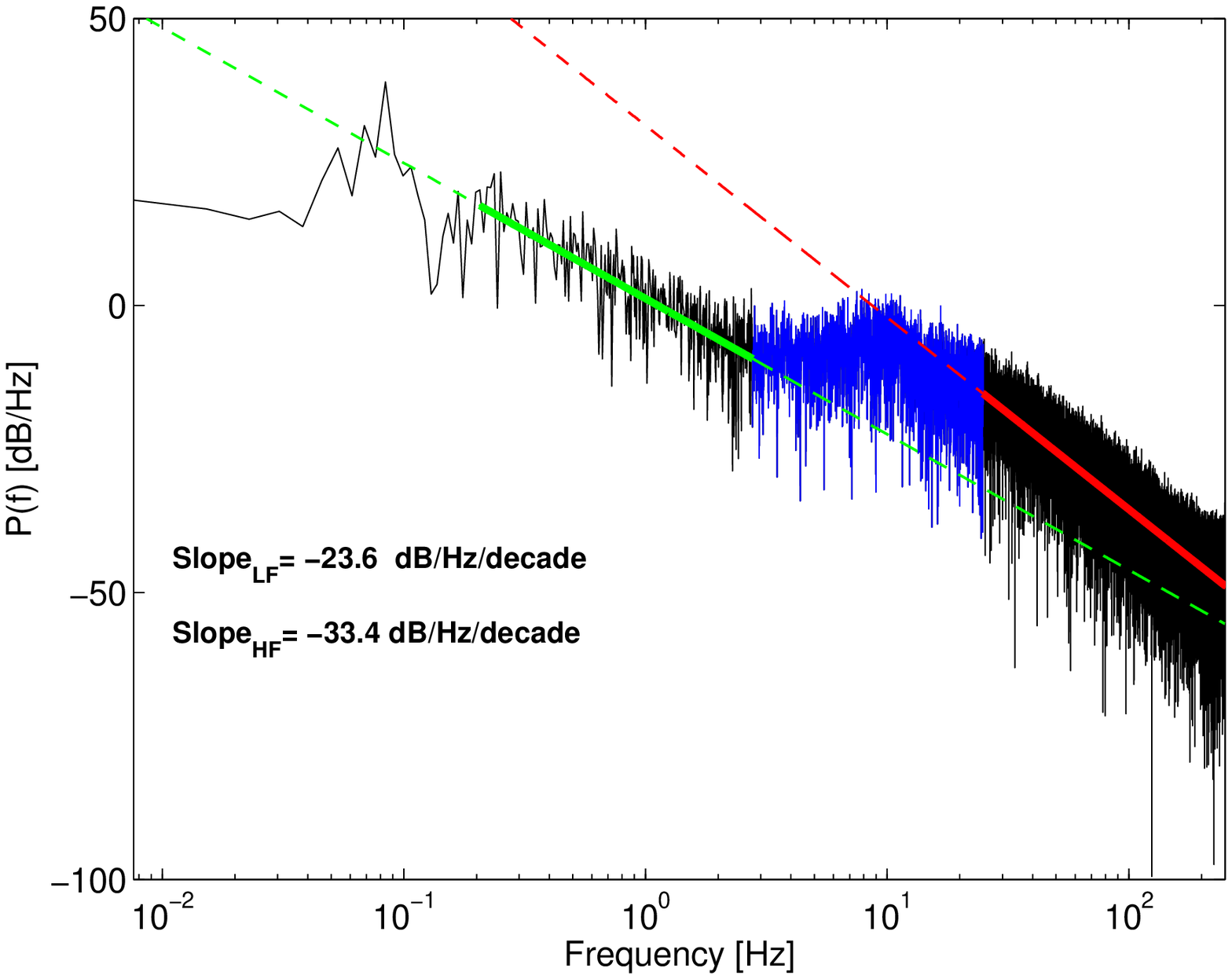}
\label{fig:trasf-e}} \hspace{2mm}
\subfloat[]{\includegraphics[scale=0.2]{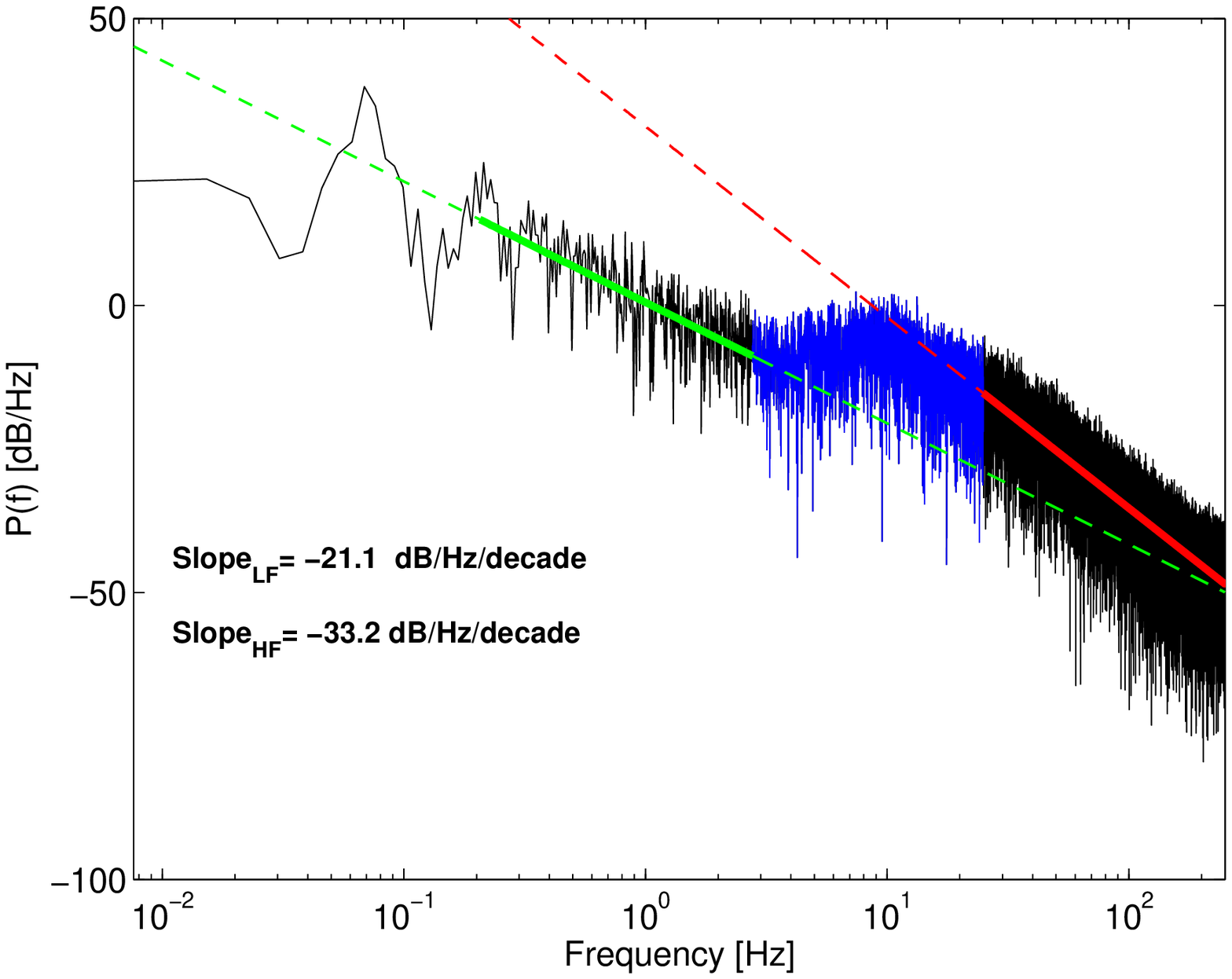}
\label{fig:trasf-f}} \\
\subfloat[]{\includegraphics[scale=0.2]{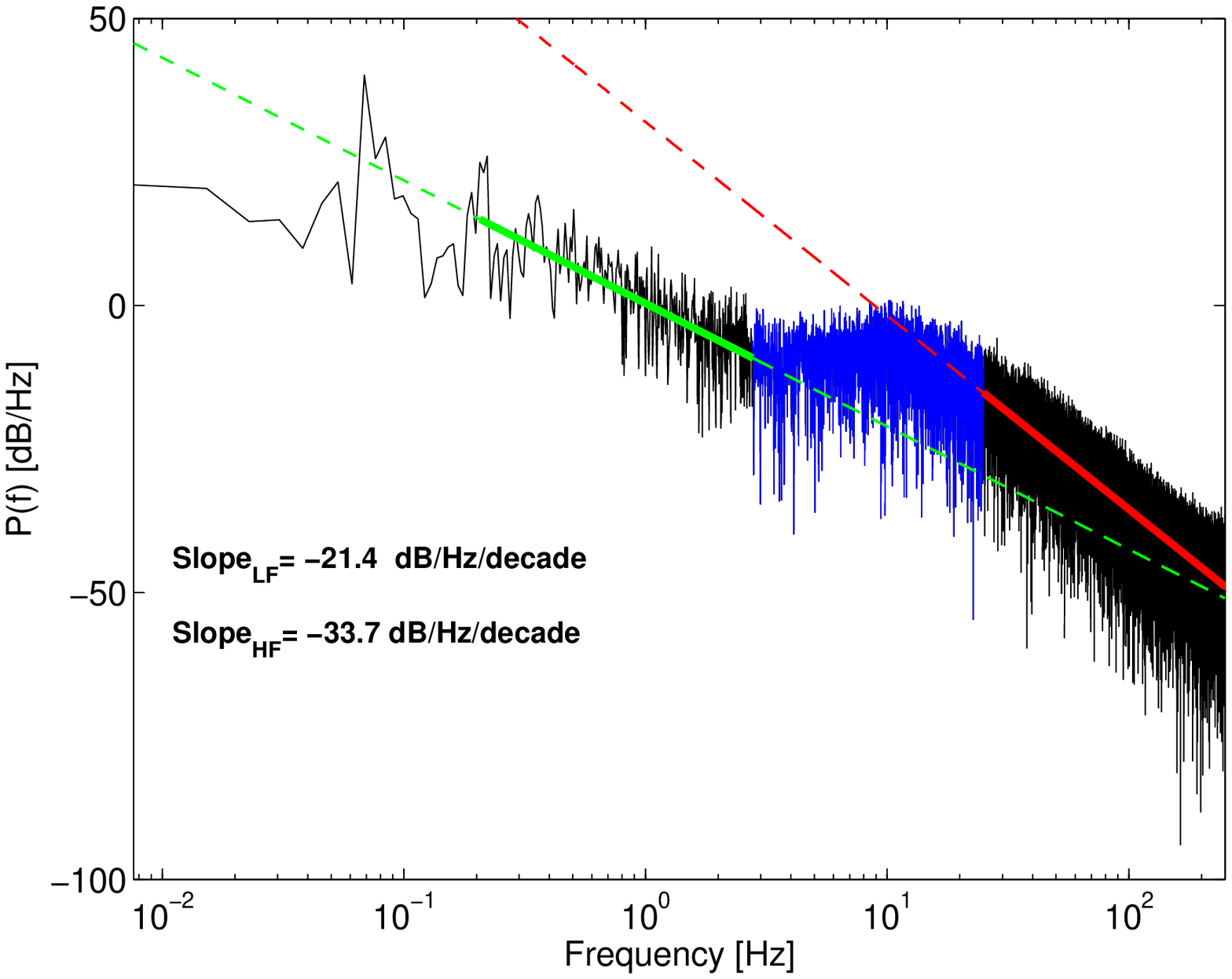}
\label{fig:trasf-g}} \hspace{2mm}
\subfloat[]{\includegraphics[scale=0.2]{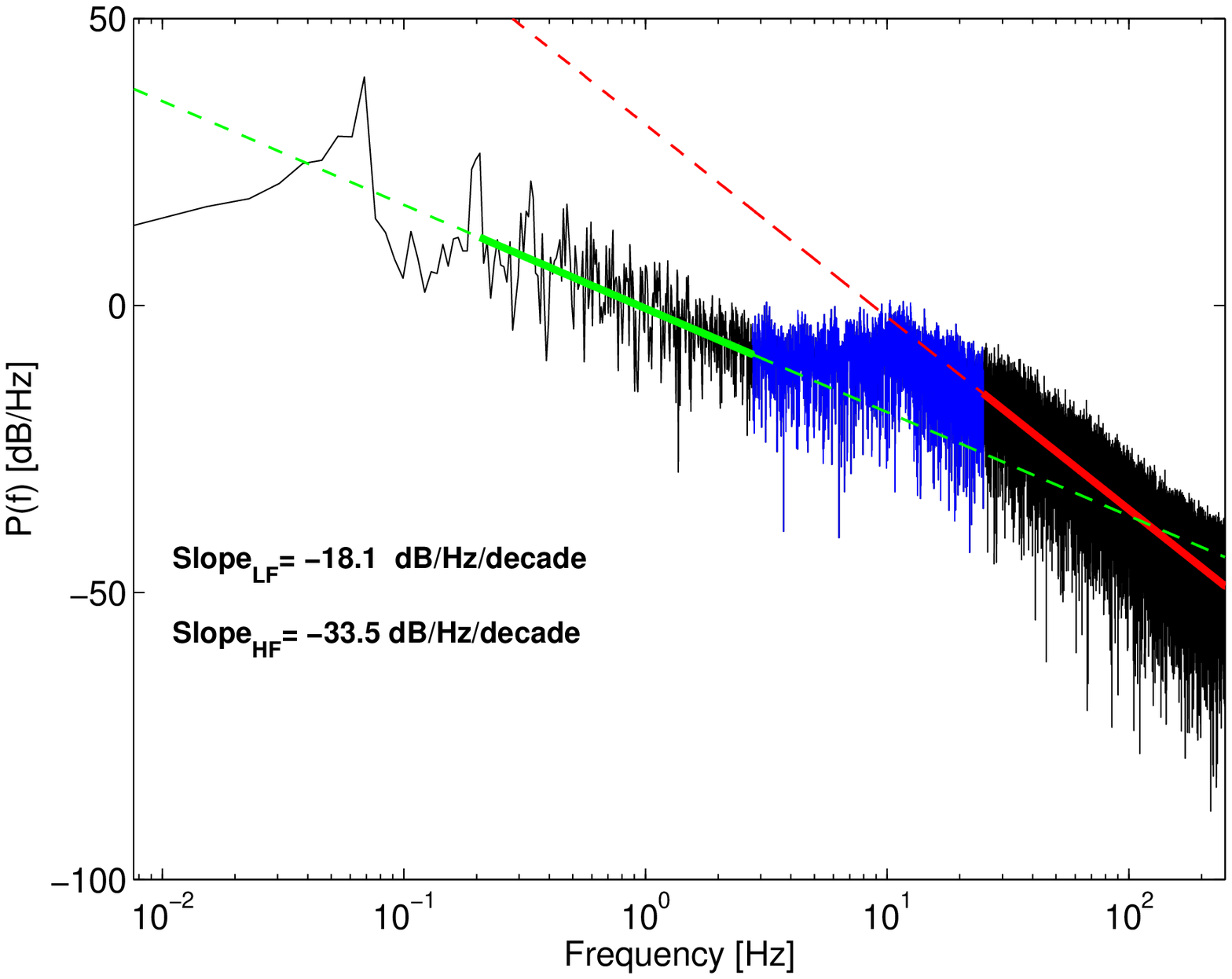}
\label{fig:trasf-h}} \hspace{2mm}
\subfloat[]{\includegraphics[scale=0.2]{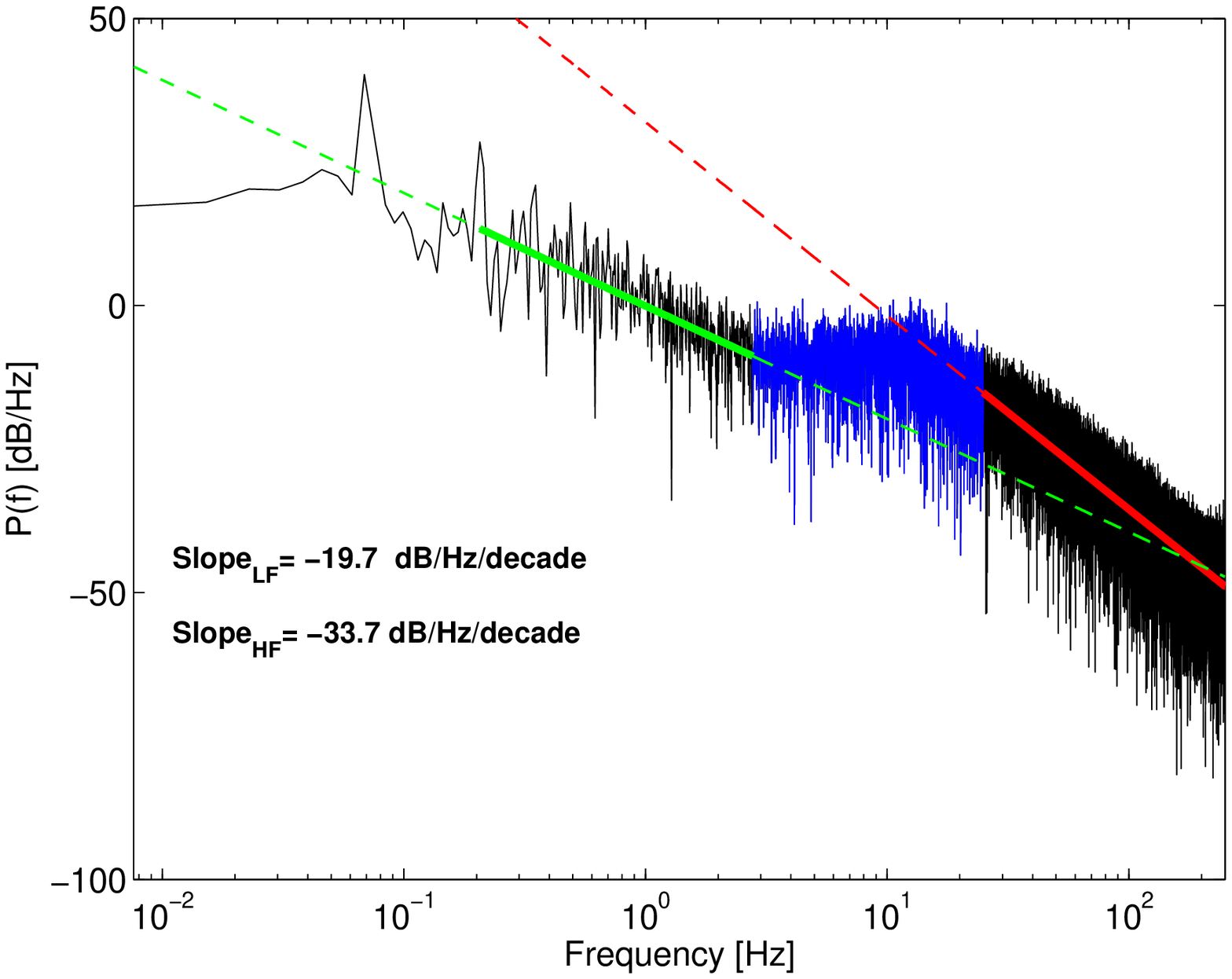}
\label{fig:trasf-i}} \\
\subfloat[]{\includegraphics[scale=0.2]{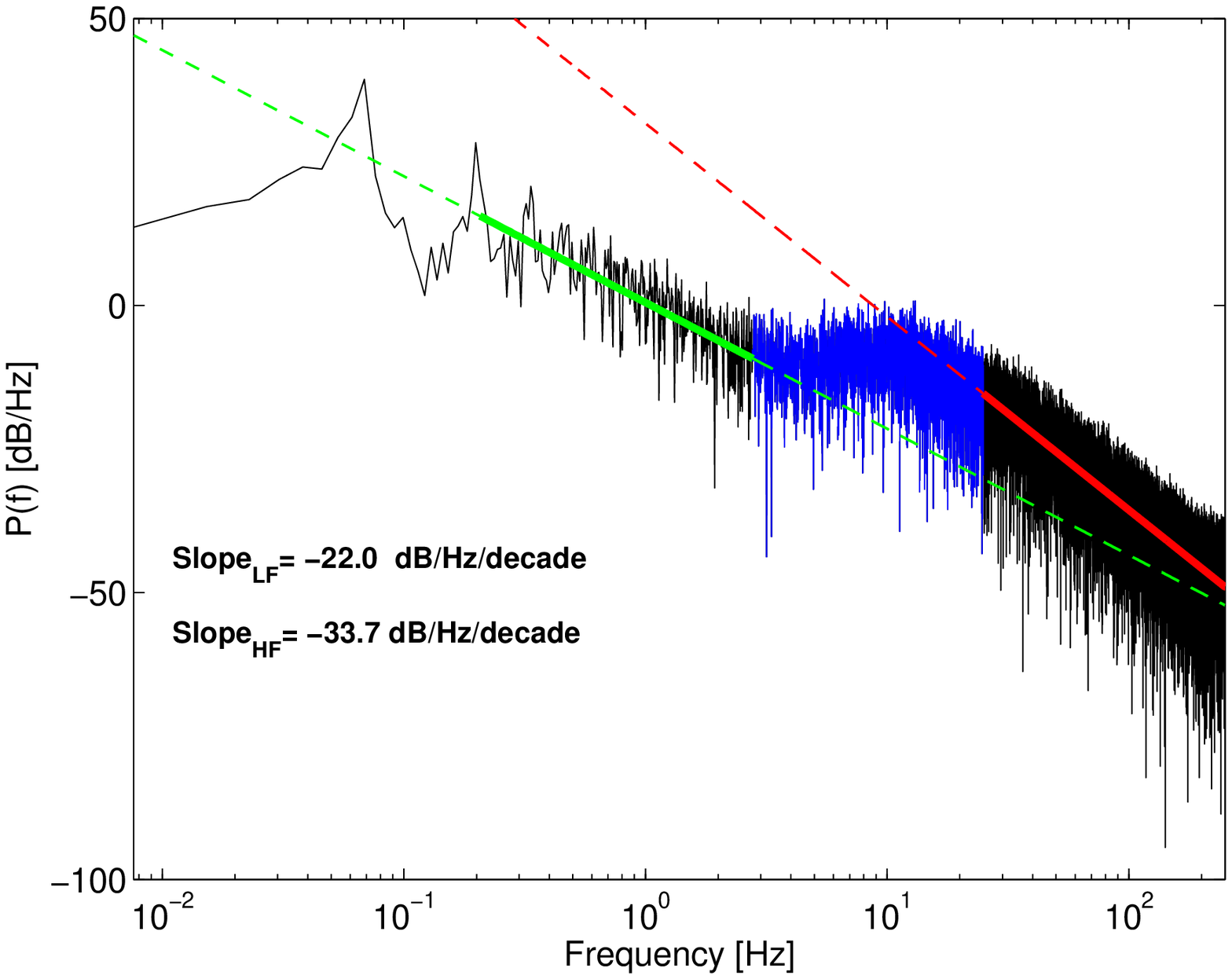}
\label{fig:trasf-l}}
\caption{Log-log power density spectra of membrane potential signals of a representative cell in clusters of increasing size, stimulated by a postprandial glucose concentration ($[G]  = 9.5\,mM$): (a) single cell; (j) $10 \times 10 \times 10$ cluster; (b-i) intermediate cases $n \times n \times n$ (with $n$ integer and $1<n<10$).
The slope at low frequencies ($S_{LF}$) is highlighted in green, the slope at high frequencies ($S_{HF}$) in red. As in Fig.~\ref{fig:tr} continuous lines segments highlight the PDS points used for the linear fitting; dotted lines segments are the extrapolation of the linear estimation. The transition region between the two linear zones is highlighted in blue.}
\label{fig:trasf}
\end{figure}

In the previous Sections, we have found that power density spectra exhibit log-log linear behavior in function of frequency with different slopes in different frequency intervals (cf. Figs.~\ref{fig:tr} and \ref{fig:trasf}). In this Section we show that an isomorphism exists between the log-log linear behavior and the coherent state formalism, which confirms the experimental observation of coherent cellular activity in the clusters.

Let $Log P$ and $Log f$ denote the ordinate and the abscissa, respectively,  in the Figures \ref{fig:tr} and \ref{fig:trasf}. Also, in each of the cases considered in the figures, let  a convenient translation of the reference frame be done, so that the straight line crosses the origin of the axis, which is equivalent to divide the abscissa  (or multiply the ordinate) by a real number $c$, with $Log \, c = s$, where $s$ is the point intercepted by the line on the abscissa axis. A measure of  self-similarity is provided by the value of the slope $d$ of the fitting straight line, $Log P = d \, Log f$, in each of the frequency intervals. Thus, we have
\be \label{pfline}
d = \frac{Log P}{Log f}~.
\ee
Since  the ratio $d$ in this equation does not depend on the logarithm base we may switch to the natural base. We will also use the notation $\al \equiv P$ and $q \equiv 1/f^d$. Equation~(\ref{pfline}) is equivalent to
\be \label{pflinen}
 u_{n,q}(\alpha) \equiv  (q \,\al)^n = 1\,, ~~~~{\rm for}~~ n\,= 0, 1,2,3,...,
 ~~~{\rm with} ~q \equiv \frac{1}{f^d},
\ee
where the frequency $f$ belongs to a given considered interval of linear behavior in the considered plots. Equation~(\ref{pflinen}) is the self-similarity relation aimed. In each of the linearity intervals, the constancy of the angular coefficient $d$ and its independence of $n$ express the scale free nature and the self-similarity properties of the phenomenon. Indeed, $d$ is called the self-similarity or fractal dimension~\cite{Peitgen} and $q$ is called the deformation or squeezing parameter~\cite{Yuen:1976vy,PLA2012,Systems2014}. We recall that, self-similarity is properly defined only in the $n \rar \infty$ limit (cf. Eq.~(\ref{pflinen})) and it expresses therefore the highly nonlinear dynamical property of the system under study (confusion should not be made between the linearity of the log-log plot and the nonlinearity of the power law Eq.~(\ref{pflinen})). The results in Figs.~\ref{fig:tr} and \ref{fig:trasf}, therefore, show that different values of $d$ in different frequency linearity intervals correspond to different nonlinear dynamical regimes of the system under study arising from different values of the control parameters such as glucose concentrations and/or cluster size.

The analysis presented in the previous Sections, based on the ODE model of the cell membrane potential $V$, stochastic gating K-Ca channels and numerical simulations is now complemented with the algebraic methods analysis proper of coherent state formalism~\cite{Perelomov:1986tf}. We observe that the functions $u_{n,q}(\alpha)$
in Eq.~(\ref{pflinen}) are, apart the
normalization factor $1/\sqrt{n!}$, the restriction to real
$q\,\al$ of the entire analytic functions in the complex $\al$-plane
\be \lab{un} {\tilde u}_{n,q}(\alpha) = {(q\, \al)^n \over \sqrt{n!}} ~,~\quad
\quad \quad~~ n\,= 0, 1,2,3,...
\ee
Thus, the self-similarity implied by our result can be studied in the space ${\cal F}$  of the entire analytic functions, by restricting, at the end, the conclusions to real $q\, \al$, $q \,\al \rar {\it Re}(q\,\al)$~\cite{QI2009,NewMat2008}. As well known, the entire analytic functions provide the key tool for the construction of (Glauber) coherent states $|\alpha \rangle$~\cite{Perelomov:1986tf,Klauder,DifettiBook}.

To see that a mathematical isomorphism indeed exists between the observed self-similarity properties and the ($q$-deformed) coherent states, we note
that ${\cal F}$ provides the
representation of the Weyl--Heisenberg
algebra of elements $\{a, a^\dag, 1\}$ with number operator $N = a^{\dagger} a$ and the Fock-Bargmann representation (FBR)~\cite{Perelomov:1986tf,Klauder}  of
the coherent states.
More explicitly, the FBR is the
Hilbert space ${\cal K}$ generated by the basis ${\tilde u}_{n}(\alpha) \equiv {\tilde u}_{n,q}(\alpha)|_{q=1}$, i.e. the
space ${\cal F}$ of entire analytic functions. A one-to-one correspondence
exists between any normalized vector
$\displaystyle{|\psi \rangle}$ in ${\cal K}$ and a function $\psi (\al) \in {\cal F}$. The vector $\displaystyle{|\psi \rangle}$
is then described by the set $\{c_n ; ~c_n~ {\rm complex ~numbers};
~\sum_{n=0}^\infty |c_n|^2 = 1 \}$ defined by its expansion in the
complete orthonormal set of eigenkets $\{ |n\rangle \}$ of $N$:
\bea \lab{psi} |\psi \rangle  = \sum_{n=0}^\infty c_n |n\rangle
&\rightarrow&  \psi (\al) = \sum_{n=0}^\infty c_n {\tilde u}_{n}(\alpha), \\
\lab{psi2} |n\rangle  &=& \frac{1}{\sqrt{n!}} (a^\dag)^{n}| 0
\rangle ~, \eea
where $|0\rangle$ is the ground state vector, $a |0\rangle = 0$,
$\langle 0|0 \rangle = 1$.
Putting $q = e^\zeta \;$, with $\zeta$ a complex number,
the $q$-deformed coherent state $|q \al \rangle$ is obtained by applying $q^{N}$~\cite{CeleghDeMart:1995}, called {\it the fractal
operator}~\cite{NewMat2008,QI2009}, to $|\al \rangle$
\be \lab{qN}  q^{N} |\al \rangle = |q \al \rangle =
\exp\biggl(-{{|q\alpha|^2}\over 2}\biggr) \sum_{n=0}^\infty \frac{(q
\alpha)^{n}}{\sqrt{n!}}~ |n\rangle ~. \ee
Note that $|q \al \rangle$ is actually a squeezed coherent state~\cite{CeleghDeMart:1995,Yuen:1976vy} with $\zeta = \ln q $  the
squeezing parameter. Thus, $q^N$ acts in ~${\cal F}$ as the squeezing operator. Notice the high nonlinearity of $q^N$ and $|q \al \rangle$.
As observed in Refs.~\cite{PLA2012,Systems2014,NewMat2008,QI2009}, the $n$-th power component $u_{n,q} (\al)$ of the self-similarity Eq.~(\ref{pflinen}) is obtained by
applying $(a)^{n}$ to $|q \al \rangle$ and restricting to real $q
\al$
\be \lab{nstage}  \langle q \al | (a)^{n} |q \al \rangle =  (q
\alpha)^{n} = u_{n,q} (\al), ~~ \qquad q \al \rar {\it Re} (q \al). \ee

Equations~(\ref{qN}) and (\ref{nstage}) are the aimed result. They indeed establish the one-to-one formal correspondence between the $n$-th power component $u_{n,q} (\al)$
with $n =
0,1,2,..,\infty$, Eq.~(\ref{pflinen}), and
the $n$-th term in the $q$-deformed coherent state series. In other words, they  provide the isomorphism relation between self-similarity properties of the observed phenomenology and the $q$-deformed algebra of the squeezed coherent states~\cite{PLA2012,Systems2014,NewMat2008,QI2009}.

In conclusion, the cellular activity of beta cells analyzed in the previous Sections can be described in terms of coherent state formalism. The low frequency (long wavelength) behavior which has been identified  as a signal of collective effect due to the islet cell population finds then a description in terms of long range cellular correlation modes.

\section{Conclusions}
\label{sec:concl}
As mentioned in Section~\ref{sec:Intro}, experimental measurements~\cite{Roc,San,Ped} show that the electrical activity of beta cells is synchronized over the islet provided convenient conditions on the glucose concentrations and islet size are satisfied. It has been also observed~\cite{She,Sher}  that coupled cells may share the entire population of K-Ca channels~\cite{She,Sher} ("channel sharing", cf. Section~\ref{sec:Math}), and thus are able to overcome noise and give rise to bursting behavior with a characteristic electrical pattern observed experimentally in beta cells within the islet. Motivated by these experimental observations, and by resorting to an extended version of the stochastic multi-cell SRK model~\cite{She}, of Hodgkin-Huxley type, we have studied the bursting activity of beta cells cubic clusters of variable size for different glycemic levels. Our  analysis has been based on such ODE model of the membrane potential $V$ with the stochastic gating of K-Ca channels and numerical simulations to reproduce the activity of single cells and clusters of variable size have been performed. In particular, we have studied the power density spectra in function of the frequency in terms of log-log plots and reached the following results. We have evidenced a linear log-log behavior with different slopes in different frequency intervals (cf. Figs.~\ref{fig:tr} and \ref{fig:trasf}). We have shown then that such a behavior highlights that scale free (power law) dynamics with self-similarity properties works at a basic level of the system components, in each of the different dynamical regimes identified by the different slopes. Finally, we have shown the existence of an isomorphism of the self-similarity properties of such a behavior with squeezed coherent states, with different slopes corresponding to different values of the squeezing parameter. The changeable working conditions used in the numerical simulations, such as the modulation of glycemic level  and different cluster sizes thus appear to drive the formation/modulation/disappearance of coherent structures at a microscopic level. One interesting consequence of our results concerns the ``robustness'' of coherent activity in each of the frequency intervals once the log-log linearity is established. As numerical simulations (cf. the comment at the end of Section~\ref{sec:sim}) and inspection of the values reported in Tab.\ref{tab:trasf} show, fluctuations in the membrane potential $V$ leave almost unaffected the value of the slope, which is consistently reflected in the coherent state dynamical stability at a given squeezing parameter value. Moreover, a critical transition occurs at the point in the plots where slope changes. This is consistent with the criticality of the variations of the squeezing parameter value~\cite{DifettiBook}. Namely, by expressing such a scenario in the jargon of many-body physics, we might say that as the number of cells in cluster increases (up to thousand in our numerical simulation) we are in the presence of a typical field theory ``phase transition'' phenomenon triggered by changes in the concentration of the glucose bath in which cell clusters are embedded. We recall indeed that squeezed coherent states with different squeezing parameters (different slopes in Figs.~\ref{fig:tr} and \ref{fig:trasf}) belong to different dynamical regimes (phases)~\cite{CeleghDeMart:1995}. It is interesting to observe that the linear behavior in log-log PDS/frequency plots in specific intervals of frequencies found in the present system, also appears in brain studies where assemblies (clusters) of large number of neurons with phase locked amplitude modulated oscillations and self-similarity properties are observed as well~\cite{Braitenberg,Linkenkaer,Hwa,Wang,Freeman2005c,PlenzThiagarajan2007,NewMat2008}. Our expectation is that, in addition to its  specific biochemical and cellular composition, coherent dynamics and self-similarity properties would play a crucial role in the evolution of a biological system through non-equilibrium processes driven by external inputs and boundary conditions.

\section*{Acknowledgements}
C.C., S.F., A.G. and A.L. acknowledge the International Center for Relativistic Astrophysics Network, ICRANet, the GNFM-INdAM; G.V and A.C acknowledge INFN for support.

\section*{Appendix}
\label{sec:App}

Here a full list of model's parameters and adopted values is given:

\vspace*{5mm}

\begin{tabular}{p{4cm} p{4cm} p{4cm}}
\hline
\rule{0pt}{3ex}{\bf Parameter} & {\bf Dimensional unit} & {\bf Value}\\
\hline
\rule{0pt}{3ex}$V_K$ & $mV$ & -75\\
$V_{Ca}$ & $mV$ & 110\\
$\bar{g}_K$ & $pS$ & 2500\\
$\bar{g}_{Ca}$ & $pS$ & 1400\\
$\bar{g}_{K-Ca}$ & $pS$ & 30000\\
$g_c$ & $pS$ & 215\\
$V_n$ & $mV$ & -15\\
$s_n$ & $mV$ & 5.6\\
$V_m$ & $mV$ & 4\\
$s_m$ & $mV$ & 14\\
$V_h$ & $mV$ & -10\\
$s_h$ & $mV$ & 10\\
$a$ & $mV$ & 65\\
$b$ & $mV$ & 20\\
$c$ & $ms$ & 60\\
$\bar{V}$ & $mV$ & -75\\
$\lambda$ & - & 1.7\\
$K_d$ & $\mu M$ & 100\\
$f$ & - & 0.001\\
$\alpha$ & $mmol\,C^{-1}\,\mu m^{-3}$ & $4.506\cdot10^{-6}$\\
$\tau_c$ & $ms$ & 1000\\
$A$ & $ms^{-1}\,mM^{-1}$ & $6.3\cdot10^{-3}$\\
$B$ & $ms^{-1}$ & 0.0147\\ [1ex]
\hline
\end{tabular}

\vspace{1cm}


\newpage

\begin{table}[h]
\caption{Computed slopes for different values of $V_t$ and increasing cells number.}
\label{tab:trasf}
\begin{center}
\begin{tabular}{p{4cm} p{2cm} p{4cm} p{4cm}}
\hline
\rule{0pt}{3ex}{\bf Number of cells} & {\bf $V_t\,[mV]$} & {\bf $S_{LF}\,[dB/Hz\,\mathrm{per\,decade}]$} & {\bf $S_{HF}\,[dB/Hz\,\mathrm{per\,decade}]$}\\
\hline
\rule{0pt}{3ex}1 & -40 & -2.1 & -38.8\\
& -20 & -1.9 & -38.7\\
& -10 & -0.1 & -39.0\\
& 0 & -1.6 & -38.7\\
& +10 & -2.8 & -39.0\\
& +20 & -2.1 & -39.2\\
& +40 & -3.5 & -38.8\\ [1ex]
\hline
\rule{0pt}{3ex}27 & -40 & -19.2  & -33.3\\
& -20 & -19.4 & -33.6\\
& -10 & -18.7 & -33.6\\
& 0 & -19.3 & -33.6\\
& +10 & -19.9 & -33.5\\
& +20 & -18.5 & -33.3\\
& +40 & -19.0 & -34.0\\ [1ex]
\hline
\rule{0pt}{3ex}64 & -40 & -19.7 & -34.1\\
& -20 & -22.3 & -33.8\\
& -10 & -20.9 & -33.9\\
& 0 & -20.2 & -33.8\\
& +10 & -21.9 & -34.0\\
& +20 & -21.3 & -33.2\\
& +40 & -21.6 & -34.0\\ [1ex]
\hline
\rule{0pt}{3ex}125 & -40 & -23.4 & -34.2\\
& -20 & -23.6 & -34.2\\
& -10 & -22.7 & -34.2\\
& 0 & -23.0 & -33.6\\
& +10 & -22.3 & -33.8\\
& +20 & -20.3 & -33.4\\
& +40 & -23.4 & -34.0\\ [1ex]
\hline
\end{tabular}
\end{center}
\end{table}


\end{document}